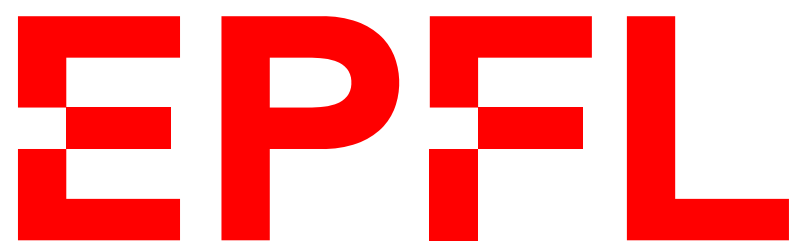

École Polytechnique Fédérale de Lausanne

# Source-to-Source Transformations for GPU Code Generation

Julien de Castelnau

Master's Thesis

Approved by the Examining Committee:

Prof. Clément Pit-Claudel
Thesis Advisor

Dr. Thomas Kœhler
External Expert, Internship Supervisor

Dr. Arthur Charguéraud
Internship Supervisor

March 20, 2026

# Acknowledgements

My advisor, Prof. Clément Pit-Claudel, has been a phenomenal mentor and teacher from the very beginning to the end of my time at EPFL, and is a major source of my inspiration to pursue a career in research. Between your mentorship during my MSc Research Scholars project, this thesis, and for making this internship possible in the first place, I cannot thank you enough.

I would also like to express my sincerest gratitude for my supervisors during this internship, Dr. Thomas Kœhler and Dr. Arthur Charguéraud, who have both invested significant amounts of time not only into the success of my project, but also into my personal development as a researcher. Thank you both for your continued advocacy and support, especially during difficult circumstances for me.

To all the other mentors I have had the joy of working with at EPFL: Prof. Giovanni De Micheli, Dr. Mingfei Yu, Prof. Thomas Bourgeat, Prof. Paolo Ienne, and Mohamed Shahawy; thank you for all that you've taught me, and for always furthering my excitement for research.

To my friends Khurshed, Cem, and Srushti, I am lucky to have met you at EPFL and am deeply grateful to have you all in my life. My fondest memories of Lausanne are times we spent together, and your presence during the most challenging times at EPFL means so much to me.

Most of all, I would not be here without the immense support of my family. My mother and stepfather have done so much to support me throughout my education, and my sister and brother-in-law have always been there for me when I have struggled the most. Thank you for everything.

*March 20, 2026* Julien de Castelnau

# Abstract

GPUs have become essential in modern high performance computing, but programming them correctly remains a significant challenge. This difficulty arises from subtle concurrency bugs that result from the explicit management of synchronization primitives and data movement across intricate hierarchies of memory and parallel threads. At the same time, the ability to control these aspects explicitly is at the core of the performance gains granted by GPUs. These challenges have motivated interest in safe GPU programming: tools and languages that can prevent or detect such bugs statically. However, existing approaches make tradeoffs in three dimensions: the strength of their guarantees, the degree of low-level control they allow, and the amount of additional effort required to achieve these safety guarantees. This thesis presents OptiGPU, a system for GPU programming with strong safety guarantees—data race freedom, deadlock freedom, and full functional correctness—that minimizes tradeoffs in the other two dimensions compared to previous approaches. OptiGPU applies proof-preserving compilation to GPU programming, allowing verification of low-level, optimized GPU programs through refinement of simple, verified CPU programs. An OptiGPU user thus avoids the substantial proof burden of directly verifying complex optimized GPU programs, instead directing this refinement with source-to-source transformations that automatically preserve proofs. OptiGPU is implemented as a set of extensions to OptiTrust, an existing framework for proof-preserving compilation on CPUs. OptiGPU models essential GPU programming features, including kernel launches, shared memory, and synchronous barriers, and produces both device and host-side code. We evaluate OptiGPU on two case studies, matrix transpose and tree-based parallel reduction, showing it can derive CUDA code matching techniques found in handwritten references.

# Contents

# 1 Introduction

GPUs have become essential for modern computing, offering massive parallelism that accelerates workloads in machine learning, scientific simulation, and graphics. However, programming GPUs effectively remains challenging. The canonical GPU programming language, CUDA, exposes a low-level programming model that gives fine-grained control over performance, but leaves the user prone to writing data races, deadlocks, and functional correctness bugs. These issues stem from the need to explicitly manage complex hierarchies of concurrent threads, synchronization primitives, and the manual orchestration of data movement between various memories.

On CPUs, runtime techniques such as bounds checking, garbage collection, and lock-order enforcement help prevent entire classes of bugs, but these are less practical on GPUs due to the emphasis on extracting maximum performance from the hardware. Similarly, while safe programming languages and static analysis tools exist for CPU code, they are designed around CPU threading and memory semantics and do not translate to the fundamentally different thread and memory architecture of GPUs. This has motivated growing interest in *safe GPU programming*: the development of tools that can prevent or detect bugs specific to GPU programs.

These tools can be broadly divided into two categories: static analyzers and programming languages. Static analyzers can catch bugs in CUDA programs, but they are generally incomplete and best-effort in nature; there may be input programs the tools cannot handle and do not provide useful results for, giving false positives or negatives. [5, 14]. Programming languages, on the other hand, can provide strong guarantees that certain classes of bugs cannot occur by construction, even if those guarantees only cover a subset of possible problems such as data races or deadlocks. In this thesis, we focus on programming languages.

Within this space, we can further distinguish between languages that rule out specific classes of bugs—such as data races and deadlocks —and those that additionally guarantee *functional correctness*, i.e., that the program computes the right result. The goal of this thesis is to achieve the latter: to provide all of these guarantees, including functional correctness, while decreasing the manual effort required from the user as compared to the state of the art. To situate this goal, we survey three existing approaches to safe GPU programming languages: high-level functional languages, low-level and safe imperative languages, and GPU verification systems.

*High-level functional languages.* Languages like Halide [19], RISE/ELEVATE [10], and TVM [9] support high-level, pure, functional array-manipulating programming with compilation to GPUs. These languages are safe by virtue of abstracting away the details of the low-level imperative code with a high-level functional model. While they may offer control over when and where values are computed and stored through scheduling directives, they all rely on a black-box imperative code generation pass that the user cannot control. These passes make performance-critical implementation decisions in the generated code, such as instruction selection, barrier placement, or how indices and loop ranges are computed. In addition, because the user cannot control the imperative code generation process, they cannot easily extend it to support new high-level constructs, new scheduling primitives, or new low-level imperative constructs that might be required for new optimizations or hardware targets.

*Low-level, safe imperative languages.* An alternative approach is to write low-level imperative GPU code directly, but in a language whose type system enforces safety properties. Languages like Descend [13] and Prism [3] take this approach, employing type systems to ensure data race freedom and deadlock freedom. As imperative languages, they offer a straightforward mapping to CUDA, meaning a user has more control over performance-critical implementation choices. Adding new imperative constructs to support new GPU features is also more straightforward. However, the requirement to statically enforce safety properties with only type annotations as input from the user often restricts their expressivity. In Descend, for example, only a predefined set of parallel access patterns on arrays (contiguous by thread, reverse, or strided, etc.) can be verified as memory safe. Accessing arrays in parallel without one of these patterns requires using an *unsafe* directive to silence the typechecker. Additionally, in Prism, barrier placement is optimized by the code generator but ultimately not controlled by the user. Regardless, these languages only guarantee basic safety properties, and do not allow proofs of functional correctness.

*GPU verification systems.* To go beyond safety properties and also guarantee functional correctness, GPU verification systems allow users to prove that their GPU code computes the correct result. Of these, to our knowledge, only Kuiper [15] models a sufficient number of GPU features—kernel launching, barriers, shared memory—to reproduce real-world CUDA programs. Kuiper offers a separation logic-based framework for verifying low-level GPU code that resembles CUDA even more closely than the safe imperative languages, as it does not need to introduce abstractions to make a fully automatic safety check possible. Crucially, Kuiper can express any optimization admissible in its proof rules, rather than being limited by what an automatic static analysis can verify. Kuiper is expressive and can verify both functional correctness and data race freedom, but it does not fully model cooperative GPU instructions such as barriers and thus does not guarantee deadlock freedom. More importantly, proving a kernel correct in Kuiper requires proving the final, fully optimized program. Since optimized GPU kernels tend to be very long due to complex loop tiling schemes and other optimizations, this generates significant proof obligations for the user in the form of writing loop invariants and ghost code (proof steps embedded in the program).

Each of these approaches represents a different tradeoff between three factors: low-level control over performance optimizations, strength of safety guarantees, and user burden. High-level functional languages require no proof steps from the user and are safe by construction, but sacrifice low-level control and expressivity. Low-level safe imperative languages restore control over the generated code, but provide only basic safety guarantees like data race and deadlock freedom, and sometimes restrict expressivity to deliver these guarantees automatically. Kuiper provides the strongest guarantees, including functional correctness on highly expressive CUDA-like code, but at the cost of substantial manual proof effort.

In this thesis, we introduce OptiGPU, a GPU programming system that aims to minimize the compromises between these three factors: to achieve functional correctness and safety guarantees comparable to Kuiper, while reducing the proof burden on the user. OptiGPU is based on the existing idea of *proof-preserving compilation*, wherein a verified program is compiled, or transformed, through steps that preserve its proof of functional correctness [2, 16, 23]. Rather than requiring the user to verify an already-optimized GPU kernel, OptiGPU lets the user write and verify a simple CPU program, and then apply correctness-preserving transformations to produce an optimized GPU kernel. OptiGPU extends OptiTrust [4], an existing proof-carrying framework for CPU code, to support GPU programming. We describe how OptiTrust, and OptiGPU in turn, implements proof-carrying code through its end-to-end flow, depicted in Figure 1.1. The specific contributions of OptiGPU with respect to OptiTrust are also highlighted in green in the figure.

First, instead of writing and proving a fully optimized GPU program correct, an OptiGPU user only proves the correctness of an initial imperative (C-like) implementation of their functional specification. The logic of this proof system is based on *separation logic* [18], a standard technique for reasoning about the correctness and memory safety of concurrent, imperative programs. OptiGPU's typechecker serves as the proof verifier for this logic. The starting implementation with its user-provided proof is depicted in the leftmost box of Figure 1.1.

The user then writes a *transformation script* that applies typical scheduling optimizations (e.g., loop tiling, reordering, fusion, array hoisting) as well as rewrite steps that describe how to offload the computation to the GPU. These GPU rewrite steps, as well as the constructs in the language required to express GPU programs, are the specific contribution of OptiGPU. OptiGPU verifies the proof still holds after every step; this is what provides the final guarantee of correctness. The middle of Figure 1.1 shows examples of transformation steps.

Crucially, since the steps are proof-preserving, they automatically discharge the majority of the proof obligations that would otherwise be required for the final optimized program. For instance, consider a loop tiling transformation: assuming constraints around divisibility of loop bounds, it is clear to see why tiling a loop would not break correctness. Yet, to directly prove an optimized program with many layers of tiled loop nests, the user would be responsible for writing the invariants and proofs on each loop, creating a significant amount of burden. Proof-preserving transformations can automate this burden in many cases, but even when the automation fails, the user has an option to intervene and fix the program's proof manually by inserting code or applying additional transformations. This "escape hatch" is shown with "Additional Proofs" next to the red arrow in the middle of the transformations in Figure 1.1.

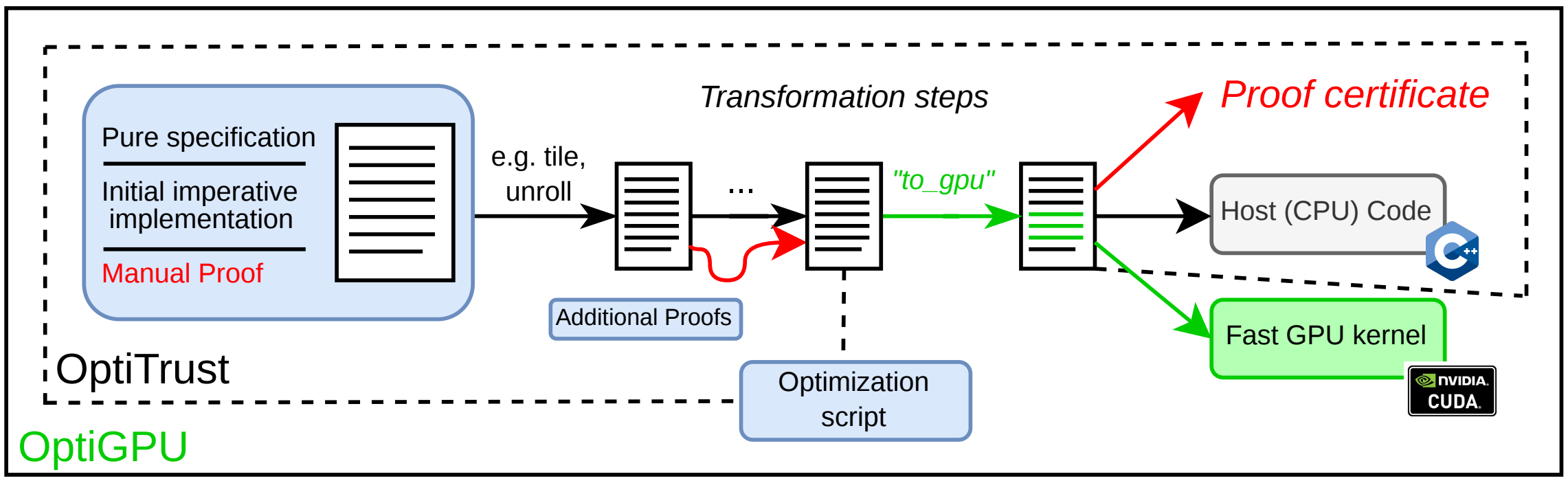

Figure 1.1: End-to-end flow of OptiGPU. Items in blue are provided by the user. The flow of the proof throughout the system is shown in red, starting from the user's input at the left and ending with the proof certificate produced for the final program after applying transformations. The parts specific to OptiGPU are highlighted in green; we highlight specific lines of the final box to denote lines of the GPU language added by OptiGPU.

The final result of applying the script to the initial program is a verified, data race free, deadlock free optimized GPU program that can be generated to CUDA. Notably, as shown by the two boxes at the right side of Figure 1.1, the produced GPU program contains both GPU "kernel" code and CPU "host" code that must accompany it to transfer data to the GPU, launch the program, and run CPU-side computations.

The contributions of this thesis are:

- A framework called OptiGPU, that enables deriving optimized and verified GPU code (and the accompanying host-side CPU code) from unoptimized, verified imperative code by applying proof-preserving transformations. Following an observation that most transformations are not specific to GPU code, we build upon the existing OptiTrust transformation framework, which previously supported only CPU code (Chapter 2 and Chapter 3).
- A language and logic to specify and verify GPU code independently of transformations, based on a collective programming and reasoning paradigm. We express this language as a set of axiomatic constructs modeling kernel launches, collections of threads, memory operations, and barriers (e.g. `__syncthreads`) in the internal language and separation logic of OptiTrust (Chapter 4).
- Two program optimization case studies demonstrating how OptiGPU enables deriving advanced CUDA code, supporting shared memory and barriers, and matching techniques leveraged by handwritten reference code such as tree-based parallel reductions. We benchmark OptiGPU generated code against a reference and quantify the lines of input required from OptiGPU users (Chapter 7).

Aside from these main contributions, we also provide details on the following technical aspects of OptiGPU's implementation:

- The proof-preserving transformations enabling the refinement of CPU to GPU code in OptiGPU (Chapter 5).
- The code generator from OptiGPU's internal GPU language to CUDA (Chapter 6).

In summary, this work applies the technique of proof-preserving compilation (starting from imperative CPU code) in the context of verified GPU programming. It supports the most essential features of CUDA, but we have not yet added support for atomics, tensor cores, asynchronous instructions, or other modern CUDA features. In addition, our case studies only consist of simpler array or tensor-like programs; we do not yet support, e.g., graph algorithms. Finally, we have axiomatized the reasoning rules of OptiGPU; we leave it to future work to present a formal semantics and a soundness proof.

# 2 OptiTrust Background

This chapter provides background on OptiTrust, an interactive source-to-source optimization framework aimed at producing verified optimized code, that is at the foundation of OptiGPU. To produce this code, OptiTrust takes as input verified and unoptimized code (Section 2.1) paired with a transformation script (Section 2.2). We will demonstrate the system by following the example of a function that computes the sum of an array (a reduction) throughout the chapter, starting with its verified input code.

## 2.1 Proof-Carrying Input Code

The input to OptiTrust is a C program with a set of *annotations* to facilitate proofs, along with a transformation script. We focus on the C program in this section. The program for our reduction example is shown in Listing 2.1.

OptiTrust translates this annotated C program to an internal language called Opti$\lambda$, an imperative lambda calculus, where both verification and transformations occur. Throughout this thesis, we will show C, or at least C-style code, but the reader should bear in mind the code presented is *not* C, and does not strictly follow C semantics. This presentation is merely a reverse translation of Opti$\lambda$ back to C (which OptiTrust also implements for the purposes of code generation).

The annotations are expressed in C using calls to functions starting with two underscores, usually with strings passed as arguments. For example, lines 2-4 in Listing 2.1 serve as a contract (specification) for the function `reduce`. These calls are no-ops from the C perspective, but inform the translation to Opti$\lambda$, where things like function contracts, loop invariants, and proof terms (also called ghosts) are first-class constructs.

C

```c
1  float reduce(float *arr, int N) {
2    __requires("A: int -> float");
3    __reads("arr ~> Matrix1(N, A)");
4    __ensures("_Res =. reduce_sum(N, A)");
5
6    float sum = 0.f;
7    __ghost(rewrite_float_linear, "inside := (fun v -> &sum ~~> v),
   by := reduce_sum_empty(A)");
8    for (int i = 0; i < N; i++) {
9      __spreserves("&sum ~> reduce_sum(i,A)");
10     sum += arr[i];
11     __ghost(in_range_bounds, "i", "i_geq_0 <- lower_bound");
12     __ghost(rewrite_float_linear, "inside := (fun v -> &sum ~~> v),
   by := reduce_sum_add_right(i, A, i_geq_0)");
13   }
14   __ghost(eq_refl_float, "reduce_sum(N, A)");
15   return sum;
16 }
```

Listing 2.1: Input C code with annotations for the reduction example.

Opti$\lambda$ itself supports a relatively standard set of operations for an imperative language, including assignment, loops, and if-then-else statements. It supports heap-allocated memory with `malloc`, `free`, etc., as well as stack memory (local variables). We note however that Opti$\lambda$ does not currently support non-terminating loops (thus all Opti$\lambda$ programs must terminate), early returns, or in general arbitrary control flow.

Setting aside these constraints, translation of the non-proof subset of Opti$\lambda$ to and from C is not the subject of this work; we refer the interested reader to [4] for a more detailed treatment of this subject. Instead, we focus on the details of OptiTrust's proof system necessary to understand OptiGPU's proof rules for GPUs. To that end, we begin our discussion with function contracts.

### 2.1.1 Function Contracts

Verification in OptiTrust is rooted in *function contracts*: specifications for the behavior of imperative functions. Contracts are described in terms of a *precondition*, the assumptions of the input arguments and input state, and a *postcondition*, the guarantees of the output values and output state. The OptiTrust typechecker is a proof checking procedure that verifies, given the precondition, after executing the function body, the postcondition holds.

These assumptions and guarantees are expressed in OptiTrust using *resources*, and they are written inside the strings for each annotation. There are two variants of resources: pure and linear.

*Pure resources* are facts that do not depend on the state of the program. Pure resources can be propositions, such as 2+2=4, a common object that proof assistants such as Rocq or Lean reason about. They can also be instances of logical types, i.e. terms without side effects, such as `f: int → float` indicating that `f` is a pure function from `int` to `float`[1].

*Linear resources* are facts about the current state of memory (in terms of the variables in the program). An example is the resource `p ~~> 1`: it says the pointer `p` points to, or holds the value 1. Linear resources derive their name from linear logic/typing: they must be used exactly once while typing a term. This is to ensure both that memory is never used after freeing, and that memory is eventually freed (no memory leaks). By contrast, pure facts like `2+2=4` cannot be "consumed", and can be used as many times as needed. Note that we also refer to linear resources as *permissions* from here.

The logic of resources in OptiTrust is a form of separation logic [18], a standard technique for reasoning about the correctness and memory safety of imperative programs, especially in concurrent settings. Linear resources in OptiTrust are equivalent to heap assertions in separation logic literature, specifically assertions extended with fractional permissions. The critical property of separation logic heap assertions is that they describe unique, disjoint parts of memory. This prevents problems that make local and modular reasoning difficult, like pointer aliasing, at the logic level.

In this thesis, we will continue to present OptiTrust's resource logic through the interface it exposes to its users in terms of annotations, rather than drawing a correspondence to separation logic literature. We refer the reader interested in OptiTrust from the perspective of separation logic to [4].

[1] In other words, pure resources are of a type that is of type `Type` or `Prop`.

To understand resources and contracts, we use the example `reduce` function in Listing 2.1. Starting on line 2, `__requires` adds a pure resource to the precondition. In this case, we are quantifying over some function `A`; this function from integer indices to floats serves as our logical model for the stateful array `arr`.[2] The permission `arr ~> Matrix1(N,A)` on line 3 captures this fact; it says that for any index `i` between `0` and `N-1` (inclusive), the `i`th element of `arr` points to the `i`th element of `A`: `&arr[i] ~~> A(i)`. In fact, `Matrix1` is syntax sugar for this "for some i" expression: `for i in 0..N -> &arr[i] ~~> A(i)`. This `for` expression is called a *group*, and is how arrays are modeled, as iterated collections of permissions on individual array elements.

The `__reads` annotation on line 3 indicates that we have a *fractional*, read-only permission on `arr`. Permissions in OptiTrust have a *fraction* from 0 to 1 attached. A fraction of 1 means a *full* permission, where one is allowed to write, free, etc. Any other fractional value means only reading is permitted. We write $\alpha$`H` to indicate the permission `H` has the fraction $\alpha$, but omit $\alpha$ if it is 1.

Permissions can be split and recombined based on their fractions; for example, a full permission `p ~~> v` can be split into two permissions of $\frac{1}{2}$`p ~~> v`, and then later recombined to form `p ~~> v` again. This system of fractions is what allows different parallel threads to share data they are only reading. We revisit this rule in Chapter 4.

`__reads(H)` is then syntax sugar for having an unspecified fraction $\alpha$ of `H` in both the pre and postcondition. In other words, the function expects some fraction of `H`, but it preserves that fraction.

Finally, on line 4, `__ensures` adds a pure resource to the postcondition. In this case, the pure resource expresses the equivalence between `_Res` – a variable name to refer to the return value of the function – with `reduce_sum(N,A)`. `reduce_sum` is a pure function that gives our functional specification: `reduce_sum(N,A)` is the sum of array `A` from 0 up to `N`. We elide its definition here, but it is given by a set of axioms that describe its properties in the base case and recursive case.

[2] Note that logical arrays are modeled as total functions in OptiTrust to simplify reasoning, but their output is unspecified outside their legal range of indices.

We also mention `__writes`, and uninitialized cells here, even though they do not appear in Listing 2.1. Upon allocation, OptiTrust assumes memory is uninitialized. Thus, instead of yielding a permission such as `p ~~> v` when `p` is allocated, it gives `p ~> UninitCell`. This permission is not sufficient to read from `p`, but it is sufficient to write to it, which would then produce a permission with the value written. `__writes(H)` in turn adds `Uninit(H)` to the precondition, and `H` to the postcondition. `Uninit` is a meta-level procedure applied by the frontend when processing these contracts to (syntactically) convert permissions like `p ~~> v` to `p ~> UninitCell`. It also applies this conversion in-depth with groups, e.g. `for i in 0..N -> &a[i] ~~> 0` becomes `for i in 0..N -> &a[i] ~> UninitCell`.

The full list of function contracts and how they affect the pure and linear parts of the preconditions and postconditions is given in Table 2.1. We also give a list of permissions and their meaning in Table 2.2.

| Contract clause | Precondition | | Postcondition | |
|---|---|---|---|---|
| | Pure | Linear | Pure | Linear |
| `__requires("x: T");` | `{x: T}` | – | – | – |
| `__ensures("x: T");` | – | – | `{x: T}` | – |
| `__consumes("y: H");` | – | `{y: H}` | – | – |
| `__produces("y: H");` | – | – | – | `{y: H}` |
| `__preserves("y: H");` | – | `{y: H}` | – | `{y: H}` |
| `__writes("y: H");` | – | `{y: Uninit(H)}` | – | `{y: H}` |
| `__reads("y: H");` | `{α: Frac}` | `{y: αH}` | – | `{y: αH}` |

Table 2.1: List of function contract annotations supported by OptiTrust. A dash (–) means the annotation does not affect the specified column. We use brackets { } as notation for a set of resources. Note that all resources can be optionally named in annotations; we have named them all here for clarity.

| Permission | Intuitive meaning |
|---|---|
| `p ~> UninitCell` | The pointer `p` can be written to but its contents are unknown. |
| `p ~~> v` | `p` can be written to and currently points to the value `v`. |
| $\alpha$`p ~~> v` where $\alpha \neq 1$ | `p` points to `v` and can only be read. |

| Permission | Intuitive meaning |
|---|---|
| `for i in 0..N -> H(i)` | The same as having the permissions `H(0)`, `H(1)`, ... `H(N-1)` in context, where `H` is parameterized by an integer `i` |

Table 2.2: List of permissions and their meaning.

### 2.1.2 Loop Contracts

Similar to function contracts, there are also *loop contracts* in OptiTrust. These are annotations that indicate the specification that the loop body must uphold for any iteration `i` of the loop. This specification then informs the effect of the loop on the outside resources. Inside the typechecker, the loop body is treated as a function of `i` and its free variables, thus loop contracts internally become function contracts. In this section, we discuss loop contracts, focusing only on the annotations necessary to understand the system, rather than a complete explanation.

First, there is the annotation `__spreserves()`, which is used in line 9 of the reduce example in Listing 2.1. `__spreserves()` expresses a *loop invariant*: a condition that must hold before the loop and at the end of every iteration. For example, in Listing 2.1, the loop invariant is that the accumulator `sum` holds `reduce_sum(i,A)`, which is the pure value for the sum of `A` up to index `i`. More specifically, `__spreserves(H)` adds `H` to the loop body's precondition, and adds `H`$[i := i.\text{next}]$ to the postcondition, where $i$ is the loop index variable and $i.\text{next}$ is the value of $i$ in the next iteration (e.g. $i + 1$). Outside the loop, it consumes $H[i := \text{range.start}]$ and produces $H[i := \text{range.end}]$, where `range.start` and `range.end` are the initial value and bound of $i$ respectively.

Second, every standard function contract annotation has a loop version with the prefix `__x`. The `x` stands for *exclusive*, because the resource inside the annotation corresponds *exclusively* to one iteration. For example, given some array `x` and loop with index `i`, the annotation `__xwrites("&x[i] ~~> i")` means that each iteration `i` consumes the permission for `&x[i] ~> UninitCell` and produces the permission `&x[i] ~~> i`. Outside the loop, groups wrap the resource inside the annotation. If the loop in this example was from `i = 0..N`, then the resource `for i in 0..N -> &x[i] ~> UninitCell` would be consumed before the loop and the resource `for i in 0..N -> &x[i] ~~> i` produced after. In essence, the exclusive loop contract annotations for linear resources entail splitting a group of some resource on the outside into one per iteration.

For pure resources, the exclusive annotations don't need to be disjoint, but are still parameterized by the index `i`. Thus, outside the loop, the resource is wrapped in a forall quantifier. For example, `__xrequires(P)` for some pure resource $P$ requires $\forall\ \text{range.start} \leq i < \text{range.stop}.P(i)$ before the loop.

We note also that the distinction between sequential invariants and exclusive annotations is how OptiTrust can easily determine if a loop is parallelizable. Any loop with state between serial iterations will have to use `__spreserves()`, whereas if all iterations can proceed in parallel, the resources for any given iteration can be exclusive.

Finally, `__sreads()` is used to split a fractional, read-only permission across iterations. Note that this is *not* the same as splitting a group of permissions (e.g. an array) to read each one separately (which would be `__xreads()`). `__sreads()` can be used with any permission, but in the case of an array, each iteration would have a read-only permission for the *entire* array. Both `__sreads()` and `__spreserves()` use an "s" to mean *shared*, because the resources are shared between iterations as opposed to being exclusive. However, unlike `__spreserves()`, a loop with `__sreads()` is parallelizable; reads to the same resource do not conflict.

Also, we note that in Listing 2.1, we read from the array `arr`, but never add the permission to the loop contract. This is because OptiTrust is able to infer certain loop contracts like `__sreads()` automatically. We also note that inside the body of the loop, we have a read-only permission of the entire array, but we are trying to read from a single cell, which requires that permission exactly. This operation on permissions is called a *focus*, and it is also something the OptiTrust typechecker has recently been extended to do automatically.

### 2.1.3 Ghosts

Ghosts are terms embedded in an OptiTrust program for the sole purpose of performing proof steps to meet the function specification. They are no-ops from the perspective of executable semantics (for Opti$\lambda$ itself, not just in C), and are erased when code is generated.

For the OptiTrust typechecker, however, they are standard function calls that simply have a void/unit return type. For example, we call the ghost function `rewrite_float_linear` on line 7 in Listing 2.1, to rewrite `0.f` in the permission `&sum ~~> 0.f` to the equivalent `sum ~~> reduce_sum(0,A)` (to meet the invariant for the loop). The operator `arg := val` that appears in the second argument to `__ghost` binds a named argument `arg` of the function to the value `val`. In this case, `inside` is a permission with a "hole" for an integer in it, expressed as a function from integer to the type for permissions. `by` is the pure fact holding the equivalence between the value in the hole before and the new value. `reduce_sum_empty` is an axiom representing the base case of the behavior of `reduce_sum`: `reduce_sum` up to index 0 of any array `A` is just 0. Calling this ghost then consumes `sum ~~> 0.f` and produces `sum ~~> reduce_sum(0,A)` after. We invoke `rewrite_float_linear` again inside the loop in a similar fashion, to complete the inductive case and satisfy the invariant.

We note that many ghosts in OptiTrust only have to be specified explicitly due to a lack of proof automation, which we acknowledge is a current limitation. For example, `eq_refl_float` (called on line 14 of Listing 2.1) just produces the fact that `reduce_sum(N,A) = reduce_sum(N,A)` (a goal that is required to satisfy the postcondition on line 4 of the program). Integrating SMT solvers, more decision procedures, and proof assistants for more complicated goals is an active line of work within OptiTrust.

Ghosts are also declared as regular functions with contracts (modulo some ghost-specific annotations for the frontend). We mention this here because ghosts are frequently *admitted* functions, which means that the OptiTrust typechecker does not verify that the function body satisfies the contract and simply assumes it is the case. This is how axiomatic proof rules can be established inside Opti$\lambda$ itself, and it is how we implement OptiGPU with minimal extensions to Opti$\lambda$'s own syntax or typechecking rules. We discuss more in Chapter 4.

## 2.2 Transformations

An OptiTrust user derives an optimized, verified program through proof-preserving, source-to-source transformations. In this section, we see how to write a *transformation script* that carries out this process, by optimizing our running example of a reduction. We also give an intuition for what it means for a transformation to be proof-preserving, and the implications for OptiTrust's implementation.

### 2.2.1 Scripts

```c
float reduce(float* arr, int N) {
  float sum = 0.f;
  float* const psums = (float*)malloc(n/128 * sizeof(float));
  #pragma omp parallel for
  for (int bi = 0; bi < n/128; bi++) {
    psums[bi] = 0.f;
    for (int i = 0; i < 128; i++) {
      psums[bi] += a[128 * bi + i];
    }
  }
  for (int bi = 0; bi < n/128; bi++) {
    sum += psums[bi];
  }
  free(psums);
  return sum;
}
```

Listing 2.2: Optimized reduce function, after transformations

```ocaml
Loop.tile (int 128) ~index:"bi" ~bound:TileDivides [cFor "i"];
Variable.local_name ~var:"sum" ~local_var:"psums"
    [tSpanSeq [cForBody "bi"]];
let factor = trm_get (trm_find_var "sum" []) in
Accesses.shift_var ~simpl:Arith.gather_rec
 ~inv:true ~factor [cFor "bi"; cVarDef "psums"];
Loop.hoist [cVarDef "psums"];
Loop.fission [tBefore; cFor "bi"; cWriteVar "sum"];
Loop.parallel [cFor "bi" ~body:[cFor "i"]];
```

Listing 2.3: Transformation script for reduce

Listing 2.2 shows the optimized version of `reduce`, which is the output after applying transformations. Note that we have omitted annotations in this program for brevity, but the transformation process maintains the annotations and produces a verified output. This program splits the work into partial sums, where the partial sums can each be computed in parallel and then summed after.

Our input to the transformations is the initial reduce from Listing 2.1. We will show the highlights of the transformation script which derives the optimized program from this point, in Listing 2.3. Transformation scripts are OCaml programs that call transformations in a sequence. Transformations, in turn, are functions with side effects on the AST. Recall that the initial C program is parsed into the internal language of OptiTrust; this is what we mean by AST. Transformations always take at least one *target* as an argument, which are lists of constraints to match against nodes in the AST. The matched nodes are then rewritten by the transformation. For example, the `cFor "i"` in the call to `Loop.tile` on line 1 targets any loop named `i` in the program. In this case, only one loop matches. Targets can also match multiple nodes; in that case, the user must be explicit about the number of matches expected.

At a high level, this script first exposes the two levels of sums by tiling the main loop. `128` is the number of elements each parallel thread will sum in the end, and `io` is the name given to the index of the new outer loop. `Variable.local_name` then creates a temporary variable `psums` (partial sums) that `sum` is substituted for, with assignments to transfer the value before and after. `Accesses.shift_var` is then used on `psums` to offset it by `-sum` (corresponding to the variable `factor` in the script) at declaration, and offset it back when it is assigned back to `sum`. After arithmetic simplification (invoked via `~simpl:Arith.gather_rec`), the declaration of `psums` becomes `psums=0`. Through these arithmetic properties, the transformations eliminate the sequential dependency of `psums` on `sum`, meaning `psums` can then be hoisted out of the loop (Line 7) into an array and each partial sum calculated independently. However, the same loop body still updates `sum`, so `Loop.fission` is used to separate the loop that calculates each partial sum from the final accumulation of all partial sums. Finally, the first loop can be parallelized using `Loop.parallel`: this sets a "parallel" flag on the loop that tells the typechecker to verify the loop is parallelizable (i.e., that it has no sequential invariant), and tells the code generator to emit a `#pragma omp parallel for` directive before the loop, as seen in Line 4 of Listing 2.2.

Transformation scripts in OptiTrust are *interactive*: the user can open a *diff* for each line of the script containing a call to a transformation, which reports the change in the AST from before to after the transformation. They can also view the diff for a series of transformations at a time. This provides better feedback to the user when trying to fix faulty transformations. Having a diff always available also allows users to focus on exactly what a transformation changed, rather than having to insert print statements on the AST in their script and read through the output. We do not discuss the interactivity of OptiTrust further as OptiGPU does not extend this aspect, but we refer the interested reader to [4].

Finally, because transformations are standard OCaml functions, they can be easily composed to build complex ones from simple ones. In fact, many of the loop transformations seen in Listing 2.2 are what OptiTrust calls *combined* transformations: rather than manipulating the AST directly, they compose so-called *basic* transformations that do. Breaking up complex transformations in this manner helps maintainability and manages the complexity of preserving proofs within transformations.

### 2.2.2 Proof-preservation

What does it mean for a transformation to be proof-preserving? OptiTrust transformations have two main burdens: updating loop contracts, and updating resources. We explain each.

Recall that OptiTrust loops carry a contract, which specifies the behavior of a single iteration of the loop . Rewriting a loop means that behavior changes, and thus maintaining the proof throughout rewrites means updating these contracts. Consider the example of applying loop fission on the simple program shown in Listing 2.4.

The original loop consists of calls to two unspecified functions `f` and `g`; each loop iteration is assumed to consume some permission `H0` and produce `H2`. The brackets on line 5 represent the current resources at that line; in this case, we assume `f()` consumed `H0` and produced `H1`. To fission this loop in between `f` and `g`, we must synthesize contracts for the new loops. This is exactly given by the resources available in between `f` and `g`, thus the first loop consumes `H0` and produces `H1`, while the second loop consumes `H1` and produces `H2`. To make this possible in OptiTrust, the typechecker exposes to the transformations the information we used here: the set of resources produced and consumed by each term. Also,

```c
1 for (int i = 0; i < N; i++) {
2   __xconsumes(H0);
3   __xproduces(H2);
4   f();
5   // { H1 }
6   // ⟵ fission here
7   g();
8 }
```

```c
1 for (int i = 0; i < N; i++) {
2   __xconsumes(H0);
3   __xproduces(H1);
4   f();
5 }
6 for (int i = 0; i < N; i++) {
7   __xconsumes(H1);
8   __xproduces(H2);
9   g();
10 }
```

Listing 2.4: Fissioning a loop and fixing its contract on a simple example. Left is before rewriting, right is after.

this simple example only uses exclusive contracts, but loops with sequential invariants may also be fissioned in certain cases, which OptiTrust implements logic for.

In addition, transformations must take care to update resources referring to variables when they are changed. Consider a variable inlining transformation. An example would be inlining `x` in `const int x = 5; y = x;` to get `y = 5;`. The transformation must make sure in this case that any resource referring to `y` like `y ⇝ x`, used in contracts or as argument to ghosts, is rewritten as well to reflect the inlining, i.e. `y ⇝ 5`.

The burden of maintaining proofs within transformations means that OptiTrust's trusted codebase is confined to its typechecker, as opposed to relying on a set of trusted transformations that perform some validation. Practically speaking, this means that not only can the user safely compose existing transformations in OptiTrust's library, but also write their own transformations that directly manipulate the AST – any unsound rewrite applied by the transformation will be caught as a broken proof by the typechecker.

## 2.3 Summary

In this chapter, we understood OptiTrust through its three major components: its language Optiλ, its resource logic (based on separation logic), and its proof-preserving transformations. We saw how users verify programs in OptiTrust, by writing annotated C code that is processed by the frontend into proof-carrying Opti$\lambda$. We then saw the rules and constructs of the resource logic that underlies these proofs. Finally, we saw how users generate verified and optimized CPU code by applying transformations on this proof-carrying input code, and briefly discussed how these transformations are implemented in OptiTrust.

OptiTrust provides benefits in each of the three dimensions of GPU programming languages that OptiGPU considers, discussed in the introduction: strength of safety guarantees, user burden, and low-level control. In terms of safety, OptiTrust gives full functional correctness as well as a foundation for reasoning about safe concurrency (lack of data races and deadlock) through separation logic. In addition, the combination of Opti$\lambda$ and OptiTrust's transformations allow low-level control over complex imperative code patterns. Lastly, proof-preserving transformations reduce user burden through automatic synthesis of loop contracts, adding ghost steps to justify transformations, and rewriting resources. OptiTrust's rich support for proof-preserving imperative code optimization make it the ideal foundation for OptiGPU to meet its goals on reducing proof effort for low-level, verified GPU programming.

In the following chapters, we will see how OptiGPU accomplishes these goals with focused extensions to each of the three major components of OptiTrust: the language (Chapter 3), logic (Chapter 4), and transformations (Chapter 5).

# 3 OptiGPU Overview

As discussed in the introduction, existing approaches to safe GPU programming force a tradeoff between low-level control, strength of safety guarantees, and user burden. OptiGPU aims to relax this tradeoff by applying proof-preserving transformations to derive optimized, verified GPU code from simple imperative starting programs.

In this chapter, we demonstrate the system by walking through the derivation of the optimized CUDA transpose kernel in Listing 3.6, starting from the CPU program in Listing 3.5. We omit proof annotations from the programs shown in this section, focusing instead on the non-proof constructs, but note that the process described here works in a proof-preserving flow.

A key observation is that most of the work can be carried out by existing CPU transformations from OptiTrust; only a small number of transformations to support refinement of CPU programs to use GPU-specific features — kernel launches, thread-parallel loops, and typed memory operations—are needed on top.

This chapter is organized in sections corresponding to each of these GPU features, starting with kernel launches.

## 3.1 Launching kernels

CUDA is a C++ dialect that supports writing GPU (called the *device*) programs, called *kernels*, as well as the CPU-side (the *host*) code to go with it. Kernels are procedures (annotated with `__global__`) that are called by the host with a special triple-chevron (`<<< ... >>>`) syntax to facilitate its *launch* (transferring the program to GPU memory and starting it) on the GPU. For example, `transpose_kernel` is launched on Line 16 in Listing 3.6. The first two

c

```c
1 void transpose(float *in, float *out, int W, int H) {
2   for (int x = 0; x < W; x++) {
3     for (int y = 0; y < H; y++) {
4       out[x][y] = in[y][x];
5     }
6   }
7 }
```

Listing 3.5: Base transpose implementation. Note that we use double brackets ($A[i][j]$) syntax sugar for accessing a 2D array.

c

```c
1 __global__ void transpose_kernel(float *in, float *out, int W, int H)
  {
2   extern __shared__ float tile[];
3
4   tile[threadIdx.y][threadIdx.x] = in[blockIdx.y * 32 + threadIdx.y *
  2 + i][blockIdx.x * 32 + threadIdx.x];
5   __syncthreads();
6   out[blockIdx.x * 32 + threadIdx.y][blockIdx.y * 32 + threadIdx.x] =
  tile[threadIdx.y][threadIdx.x];
7 }
8
9 void transpose_host(float *in, float *out, int W, int H) {
10   float *d_in, d_out;
11   size_t data_sz = sizeof(float) * width * height;
12   cudaMalloc((void **)&d_in, data_sz);
13   cudaMalloc((void **)&d_out, data_sz);
14   cudaMemcpy(d_in, in, data_sz, cudaMemcpyHostToDevice);
15
16   transpose_kernel<<<vec3(W/32,H/32),
  vec3(32,16),32*32*sizeof(float)>>>(d_in,d_out,W,H);
17
18   cudaMemcpy(d_out, out, data_sz, cudaMemcpyDeviceToHost);
19   cudaFree(d_in);
20   cudaFree(d_out);
21 }
```

Listing 3.6: Reference transpose CUDA code; the goal after applying transformations.

parameters in the triple brackets control the amount of parallel blocks and threads the kernel will run with, which we explain further in Section 3.2. The last controls the amount of bytes available for shared memory, a fast memory within the GPU with limited resources, thus needing to be pre-determined by the host. We explain shared memory and other GPU memory in Section 3.4.

OptiGPU uses a sequence of calls to intrinsic "transition" functions - `kernel_launch`, `kernel_setup_end`, `kernel_teardown_begin`, and `kernel_kill` - for kernel launches, instead of forcing device code to appear in its own function. This simplifies performing transformations across the host and device, such as manipulating certain GPU-specific memory that must be allocated and freed by the host but is accessed by the device. OptiGPU considers all code in a scope beginning with `kernel_launch` and ending with `kernel_kill` to be device code. From the typechecking perspective, GPU operations are only legal within device code. The code corresponds to the body of a CUDA kernel function, with the caveat that the other two transition functions form *setup* and *teardown* phases that represent behavior happening transparently by the CUDA runtime during the launch, not in the kernel. For example, allocating shared memory happens in the setup phase, while it is freed in the teardown phase.

In Listing 3.7, we introduce a kernel launch to our transpose program through a transformation that just wraps the target provided by the user in a scope with transition functions. The parameters to `kernel_launch` are provided by the user. These are identical to those of the triple brackets syntax, except that OptiGPU does not yet take vectors as dimensions, so the block and thread counts have been flattened.

Note that although we have indicated the program inside is device code, if we attempted to generate CUDA right now, we would get a syntax error due to the use of CPU-specific memory operations (the default) in device code. We must use transformations to make the GPU memory desired explicit, which in turns requires parallelizing the program to run on GPU threads. In the next sections, we will see how to apply each of these transformations.

c

```c
1  void transpose(float *in, float *out, int W, int H) {
2    {
3      kernel_start((W/32)*(H/32), 32 * 32, sizeof(float) * 32 * 32);
4      kernel_setup_end();
5      for (int x = 0; x < W; x++) {
6        for (int y = 0; y < H; y++) {
7          out[x][y] = in[y][x];
8        }
9      }
10     kernel_teardown_begin();
11     kernel_kill();
12   }
13 }
```

Listing 3.7: Transpose after kernel launch transformation.

## 3.2 Thread Hierarchy

A CUDA program organizes parallel execution into a hierarchy of threads, blocks, and grids. Threads are grouped into blocks, and blocks are collectively organized into a grid. A kernel launch specifies the dimensions of this hierarchy — the number of blocks in the grid and the number of threads per block — using the triple-chevron syntax (<<<...>>>) seen earlier.

Within kernel functions, CUDA provides built-in index variables `threadIdx.x` and `blockIdx.x` to identify the current thread within its block and the current block within the grid, respectively. The variables `blockDim.x` and `gridDim.x` give the number of threads per block and the number of blocks in the grid, corresponding to the parameters passed in the kernel launch configuration.

CUDA further supports multi-dimensional indexing through the `.x`, `.y`, and `.z` components of these variables, allowing the thread hierarchy to be structured in up to three dimensions. This is particularly useful when the problem domain is naturally multi-dimensional. For example, on line 4 of Listing 3.6, `blockIdx.y` and `threadIdx.y` index into the row dimension of the `in` matrix, while `blockIdx.x` and `threadIdx.x` index into the column dimension. In this way, each thread in the grid is mapped to a unique element of the matrix.

Following the design of other languages supporting safe GPU programming such as Descend [13] or TVM [9], in OptiGPU, these indices are realized into the loop indices of explicit fork/join-style parallel loops. These loops extend a standard fork/join to allow explicit barriers, which we will discuss later.

The advantage of materializing these parallel loops is that a parallel program can be easily derived through transformations by simply changing the loop “mode” of some loop nest into a “parallel” or “GPU thread” loop. The bulk of the work, deriving that loop nest in the first place, is handled by existing, GPU-unaware transformations. Compiling this loop nest to CUDA is also a matter of simply erasing the loops and mapping the loop indices to the built-in indices `threadIdx`, `blockIdx`, etc. The bodies of the loops remain as thread-level code. Lastly, this design allows inheriting OptiTrust’s existing reasoning about memory safety for fork/join parallel loops on CPUs.

To demonstrate with an example, consider the next step of the transpose program, in Listing 3.8. The original loop nest has been tiled and the result of reading `in` hoisted to expose a 32x32 temporary matrix `tile`. As indicated by the comments, the loop indices `by,bx,y,x` now match exactly with the builtin CUDA indices in Listing 3.6: this loop nest describes a serialized version of the same kernel, and the next step is to convert the loops to a parallel form.

Note the scoping of `tile` as well: its allocation happens at the loop corresponding to an individual block, aligning with the use of `tile` in Listing 3.6 as block-local *shared memory*. We discuss how OptiGPU handles the GPU memory hierarchy in Section 3.4.

## 3.3 Synchronization

CUDA introduces a set of synchronization primitives for communication across threads. For example, `__syncthreads()` acts as a barrier for all threads within a block. Without such barriers, all conflicting accesses in shared and global memory between threads are treated as undefined behavior. Crucially, the same `__syncthreads` must also be executed by all threads within the block; Listing 3.9 gives an example of an invalid program. One thread executes the first `__syncthreads` based on the conditional, while the remaining execute the

C

```c
...
for (int by = 0; by < H/32; by++) { // blockIdx.y
  for (int bx = 0; bx < W/32; bx++) { // blockIdx.x
    float* const tile = MALLOC2<float>(32,32);
    for (int y = 0; y < 32; y++) { // threadIdx.y
      for (int x = 0; x < 32; x++) { // threadIdx.x
        tile[y][x] = in[by*32+y][bx*32+x];
      }
    }
    for (int y = 0; y < 32; y++) { // threadIdx.y
      for (int x = 0; x < 32; x++) { // threadIdx.x
        out[bx*32+y][by*32+x] = tile[x][y];
      }
    }
    free(tile);
  }
}
...
```

Listing 3.8: Body of transpose after loop transformations.

other. Different threads taking different control paths is called *thread divergence*, and when barriers are in those control paths, it can lead to deadlock.

C

```c
if (threadIdx.x == 0) {
  __syncthreads();
} else {
  __syncthreads();
}
```

Listing 3.9: Undefined behavior due to thread divergence in CUDA.

Barriers, and other instructions that communicate between threads pose a problem for the simple fork/join model: each parallel iteration of the loop no longer necessarily modifies disjoint resources. While it is possible to give a thread-local reasoning rule to barriers using concurrent separation logic [7], it requires externalizing the proof that the barrier communicates resources correctly, adding more burden to the user. In addition, proving correct use of memory resources does not suffice to prevent divergent threads from executing a barrier.

C

```c
1  ...
2  thread for (int by = 0; by < H/32; by++) {
3    thread for (int bx = 0; bx < W/32; bx++) {
4      float* const tile = MALLOC2<float>(32,32);
5      thread for (int y = 0; y < 32; y++) {
6        thread for (int x = 0; x < 32; x++) {
7          tile[y][x] = in[by*32+y][bx*32+x];
8        }
9      }
10     blocksync();
11     thread for (int y = 0; y < 32; y++) {
12       thread for (int x = 0; x < 32; x++) {
13         out[bx*32+y][by*32+x] = tile[x][y];
14       }
15     }
16     free(tile);
17   }
18 }
19 ...
```

Listing 3.10: Body of transpose after `thread for` conversion.

Indeed, Kuiper, which models barriers this way, does not currently guarantee deadlock freedom from thread-divergence.

OptiGPU instead treats barriers semantically as operations executed collectively by some group of threads. A `thread for`, OptiGPU's variant of a parallel for-loop for GPUs, exposes this collective view of executing threads, by *narrowing* the view surrounding the loop. In any given `thread for`, the hierarchy of `thread for`s surrounding it describes what threads the code inside the body is running on.

For example, Listing 3.10 shows the tiled transpose from earlier, with the loops parallelized with `thread for`, and barriers inserted appropriately. Note that we present C code here, but `thread for` exists exclusively in the intermediate language of OptiGPU, and this is just a printing of that language to C with some imagined constructs.

Outside this loop nest, the code is executed by an entire grid of threads: this is then narrowed by the outer two block loops (`bx` and `by`) to one block. `__syncthreads`, which we call `blocksync` in OptiGPU, appears at this level, because synchronization is needed to facilitate the changing ownership of `tile` elements by threads. Finally, the code inside the two loop nests of `y` and `x` narrows further (the loops narrow further?) to a single thread, where memory operations must take place.

OptiGPU enforces that operations are executed by the expected threads with its extensions to OptiTrust's resource logic. Operations with thread-level side effects like loads and stores cannot appear at the block level, while barriers must appear at the block level. This prevents thread divergent barriers by construction.

OptiGPU also extends the rules over memory in `thread for` such that explicit barriers are required if and only if they are needed to prevent data races. In a fork/join, there is a joining of memory resources after the loop by default; `thread for` does not perform this join, leaving memory after the loop in a "desynchronized" state that requires a barrier to be invoked if a new `thread for` would be launched with conflicting memory accesses. We also extend OptiTrust's separation logic to reason about this desynchronized memory (Section 4.3).

Finally, as mentioned, from the perspective of the user, transforming the serial program in Listing 3.8 to Listing 3.10 only requires invoking a transformation that converts the *mode* of the loop from sequential to `thread`. Behind the scenes, these transformations implement some heuristics to insert barriers minimally. The barrier in this transpose program was inferred automatically, for instance. Note that at any time, the user can move, add, or remove barriers to the program using existing transformations from OptiTrust, as long as doing so does not break the proof of the program, or if they can apply further transformations to fix a broken proof. Thus, the automatic barrier heuristics only add convenience, and do not sacrifice user control.

We discuss the details of these transformations further in Chapter 5.

## 3.4 Memory Hierarchy

In addition to the thread hierarchy, CUDA exposes a memory hierarchy that reflects the physical architecture of the GPU. Each level trades off capacity for speed and scope of access. *Global memory* corresponds to the on-board DRAM of the GPU and is the primary means of transferring data to and from the host. It is accessible by all threads across all blocks, but is the slowest level of the hierarchy. *Shared memory* is a fast, on-chip scratchpad local to each block. CUDA programs use shared memory as a working buffer, allowing threads within a block to exchange data and reuse loaded values without repeated trips to global memory. Finally, *registers* are the fastest and most limited memory, private to each thread. Local variables in CUDA kernel functions are placed into registers unless register pressure forces them to be spilled into slower memory.

Each of these memories has rules surrounding their visibility within the thread hierarchy, and to the host. For example, none of them can be read from or written to by the host, but global memory is allocated by the host with the CUDA runtime function `cudaMalloc` (lines 12 & 13 of Listing 3.6), and copied to and from host memory with `cudaMemcpy` (lines 14 and 18 of Listing 3.6). Likewise, the GPU cannot access host side memory directly.

OptiGPU implements these rules by parameterizing standard C memory operations - load, store, allocate, free - by a *memory type*. Within code declared as device-side, only the operations on the aforementioned GPU memory types are syntactically legal. The function contracts for these operations ensure that they are only used on pointers that belong to the corresponding memory, as well as that they are executed where expected (by a single thread, a block, the host, etc.). These contracts are discussed in Section 4.3.

In the initial imperative program, all memory operations are of a generic, default type. This memory type can be directly mapped to CPU side code, but transformations are needed to refine generic memory to one of the GPU memories. Listing 3.11 shows the final state of the transpose program, after refining the memories to GPU. The intrinsic functions `__gmem_get` and `__gmem_set` perform load and store on global memory respectively; `smem` is used with similar operations for shared memory.

Before converting memory types, we have used standard (GPU-unaware) OptiTrust transformations to create local copies of the input and output arrays (`d_in` and `d_out`), and hoist the `tile` allocation to directly after the launch (vice versa for `free`). These local copies become the global memory used by the kernel, and `tile` becomes the shared memory tile corresponding to line 2 in the CUDA transpose (Listing 3.6). Note that hoisting `tile`

C

```c
1  void transpose(float *in, float *out, int W, int H) {
2    float* const d_in = __gmem_malloc2<float>(H,W);
3    memcpy_host_to_device2(d_in, in, H, W);
4    float* const d_out = __gmem_malloc2<float>(W,H);
5    {
6      kernel_start((W/32)*(H/32), 32 * 32, sizeof(float) * 32 * 32);
7      float* const tile = __smem_malloc2<float>(32,32);
8      kernel_setup_end();
9      thread for (int by = 0; by < H/32; by++) {
10       thread for (int bx = 0; bx < W/32; bx++) {
11         thread for (int y = 0; y < 32; y++) {
12           thread for (int x = 0; x < 32; x++) {
13             tile[DMINDEX(H/32,W/32,by,bx)][y][x]
14               = in[by*32+y][bx*32+x];
15           }
16         }
17         blocksync();
18         thread for (int y = 0; y < 32; y++) {
19           thread for (int x = 0; x < 32; x++) {
20             out[bx*32+y][by*32+x]
21               = tile[DMINDEX(H/32,W/32,by,bx)][x][y];
22           }
23         }
24       }
25     }
26     kernel_teardown_begin();
27     __smem_free(tile);
28     kernel_kill();
29   }
30   memcpy_host_to_device2(d_in, in, H, W);
31   gmem_free(d_out);
32   gmem_free(d_in);
33 }
```

Listing 3.11: Final OptiGPU transpose program, after converting CPU memory to GPU memories.

added extra dimensions to the matrix, `H/32` and `W/32`, that do not correspond to that of the CUDA shared memory declaration. These *distributed dimensions* are used for reasoning by OptiGPU, but are erased when generating CUDA. They must be accessed with the `DMINDEX` function, which can be thought of as mapping indices of distributed dimensions to regular dimensions. We explain distributed dimensions and `DMINDEX` further in Section 4.3.

The conversion itself is handled by a transformation that takes a target to a variable declaration (an allocation) and converts all operations on that variable to the desired memory type. If those operations appeared in the correct context to begin with - for example, the allocation for *tile* appearing in the setup portion of the kernel launch - then no extra proof steps are needed.

Having declared the kernel launch, thread hierarchy, and correct use of GPU memories, the program in Listing 3.11 is ready to be compiled to CUDA, a process that we describe in Chapter 6. Note that in the printing process, intrinsic calls like `__gmem_get` are replaced with standard assignment and dereferencing as CUDA does not make a syntactic distinction between operations on the various memories.

## 3.5 Summary

In this chapter, we saw how OptiGPU enables expressive and concise verified GPU programming through the example of optimizing a matrix transposition kernel. We saw that, thanks to the central *thread for* construct in OptiGPU, the process of establishing parallel threads, using the various special GPU memories, and managing synchronization is handled in large part by existing CPU loop and memory transformations. The remaining transformations are highly GPU-specific, simple "conversions" performed at the end of the transformation script. OptiGPU avoids much of the complexity of maintaining correctness though transformations on parallel code by moving as much of the work as possible to transformations on serial CPU code.

The GPU features seen in this chapter are all made possible by extensions to the language, logic, and verification system of OptiTrust, whcih we discuss next.

# 4 Language & Logic

In this chapter, we detail the language constructs and associated typechecking rules OptiGPU adds to Opti$\lambda$ (the internal language of OptiTrust) to support GPUs. Most of the constructs and rules are implemented within Opti$\lambda$ itself. For instance, the new operations for GPU memory types are not in the syntax of the language, but are merely declared as functions with admitted contracts. Recall that an admitted function is one where the typechecker does not verify the body satisfies the contract. We use admitted functions the establish GPU operations according to an axiomatic (in our logic of resources) interpretation of their semantics. We leave formalizing an operational semantics for these constructs to future work.

As a result of OptiGPU's embedding in Opti$\lambda$, we use a slightly unconventional presentation for a language and typing rules in this chapter. We do not define a grammar, a typing judgement, or provide horizontal rules[3], but rather show the function contracts for the new constructs and ghosts in the C-like syntax for Opti$\lambda$ we have seen so far (with some modifications for brevity). OptiGPU also adds new constructs in the resource language, which we explain along the way.

The only exception to our presentation using function contracts is `thread for`, which requires syntactic extensions to Opti$\lambda$ and a built-in typechecking rule. For this, we show the rule in a more typical horizontal bar style, albeit with some simplifications to avoid getting into the minutiae of triples and the typing judgement.

Each section of this chapter covers the implementation of one of the GPU features seen in Chapter 3. We start with the implementation of kernel launches.

[3] We refer the reader interested in the formal syntax and type system of Opti$\lambda$ to [4].

OptiΛ
```
1 void kernel_launch(int bpg, int tpb, int
  smem_sz):
2 consumes HostCtx;
3 produces KernelSetupCtx;
4 produces KernelParams(bpg, tpb,
  smem_sz);
5 produces SMemAllowance(smem_sz);
```

OptiΛ
```
1 void kernel_setup_end():
2 requires bpg: int, tpb: int, smem_sz:
  int, grid_sz: int;
3 requires by: bpg * tpb = grid_sz;
4 consumes KernelSetupCtx;
5 produces ThreadsCtx(0..+grid_sz);
6 preserves KernelParams(bpg, tpb,
  smem_sz);
7  // Must allocate all shared memory
  requested
8 produces SMemAllowance(0);
```

OptiΛ
```
1 void kernel_teardown_begin():
2 requires bpg: int, tpb: int, smem_sz:
  int, grid_sz: int;
3 requires by: bpg * tpb = grid_sz;
4 consumes ThreadsCtx(0 ..+ grid_sz);
5 produces KernelTeardownCtx;
6 preserves KernelParams(bpg, tpb,
  smem_sz);
7  // Shared memory is still allocated
  when teardown begins
8 preserves SMemAllowance(0);
```

OptiΛ
```
1 void kernel_kill():
2 requires tpb: int, bpg: int, smem_sz:
  int;
3 consumes KernelParams(bpg, tpb,
  smem_sz);
4 consumes KernelTeardownCtx;
5 // Must give back all shared memory
6 consumes SMemAllowance(smem_sz);
7 produces HostCtx;
```

Figure 4.2: Contracts for kernel transition functions.

## 4.1 Kernel Transition Functions

OptiGPU represents both host (CPU) and device (GPU) code. The transition functions (`kernel_launch`, `kernel_setup_end`, `kernel_teardown_begin`, and `kernel_end`) are responsible for enforcing the proper transition to device code, which will eventually be compiled as a kernel and the associated launch (with triple chevron syntax) in CUDA.

The proof rules for each function are shown in Figure 4.2. In this simplified syntax, we only include the C function prototype, write contract annotations as keywords instead of function calls, and write the resources without strings. Unless otherwise specified, all functions shown in this chapter are admitted.

These functions make use of what we refer to as *execution contexts* — arbitrary, axiomatized permissions that indicate where code is running. When a given execution context is in the set of resources available while typechecking some term (its precondition), that term is logically thought of as executing on the corresponding device, or thread, etc.

To start, `kernel_launch` consumes the permission `HostCtx`, and produces `KernelSetupCtx`. This means the code before `kernel_launch` is expected to run on the host, and after is part of the setup phase mentioned in Section 3.1. `kernel_kill` does the inverse, only it is coming from the teardown phase (`KernelTeardownCtx`).

As OptiGPU expects the user to start with CPU (or CPU-like imperative) code, `HostCtx` is a permission the user adds to their function specification, indicating that they expect their top-level function to be called by something that can launch kernels (i.e. acts as the host in CUDA terms).

The function `kernel_launch` also produces a permission `KernelParams`, holding the launch parameters throughout the body of the kernel. These launch parameters are relevant when allocating shared memory — the amount allocated should not exceed the amount requested at launch — or when checking that a block-level barrier is in fact executed at the block level.

At the end of the setup phase, when actual GPU code begins, `KernelSetupCtx` is replaced with `ThreadsCtx`. `ThreadsCtx` is the central execution resource that allows OptiGPU to verify that instructions are executed at the expected levels, such as a barrier being executed by a block, or memory operations being executed by a single thread. It holds an interval $[t_0, t_f)$ of thread IDs that are indexed globally in the entire grid, where `ThreadsCtx` $([t_0, t_f))$ means that interval of threads is executing the current code. Note that in the code, we use the syntax `a..b` for the interval $[a, b)$, and the syntax `a..+b` for the interval $[a, a + b)$.

At the start of the kernel, i.e. right after setup, the entire grid of threads is executing the code, thus `ThreadsCtx`$([0, tpb \cdot bpg])$ is produced, where $tpb$ is the number of threads per block requested, and $bpg$ is the number of blocks per grid. Conversely, `kernel_teardown_begin` consumes the same `ThreadsCtx`.

Finally, `kernel_launch` produces the permission `SMemAllowance`, which holds the current number of bytes remaining to allocate shared memory. This is used by the proof rule for shared memory allocation and freeing to ensure that more memory than requested is not allocated. We also enforce that *all* the memory requested is allocated with `__smem_malloc`, thus the `preserves SMemAllowance(0)` in the `setup` and `teardown` contracts, although this is not strictly necessary for correctness. We discuss the shared memory allocation rules further in Section 4.3.

## 4.2 GPU Thread Loops

As explained in Section 3.2, GPU threads are captured by a parallel fork/join-inspired loop called `thread for`. For context, let us first understand the standard `for` typechecking rule in OptiTrust, shown in Listing 4.12.

```
t(i):
consumes I(i), H0(i)
produces I(i+1), H(i)
-------------------------------------------------- T-For
for (int i = 0; i < N; i++) { t(i) }:
consumes I(0), (for i in 0..N → H0(i))
produces I(N), (for i in 0..N → H(i))
```

Listing 4.12: Standard `for` loop proof rule in OptiTrust.

Here, we are stating the premise and conclusion of the proof rule as the expected precondition and postcondition for the given term, or in other words the resources before and after executing the term. This is typically represented using triples (e.g. $\{P\}C\{Q\}$), but we use a simplified syntax that more closely aligns with the contract notation we have seen so far. The term is on the first line proceeded by a colon, then the resources in `consumes` form the precondition, and the resources in `produces` form the postcondition. Viewing the rule as a whole, the top is the premise, saying if this term satisfies this precondition and postcondition, then the conclusion on the bottom says the bottom term consumes and produces the resources specified.

In this rule, `t` is the term corresponding to the body of the loop, parameterized by the loop index `i`. `H0` and `H` are the exclusive permissions in the precondition and postcondition respectively: `__xconsumes` and `__xproduces` in terms of annotations. `I` is the invariant permission (`__spreserves`). Recall from Section 2.1.2 that exclusive permissions become groups (`for i in 0..N → ...`) on the outside, while the invariant permission `I` must hold for `i=0` before, and then it holds for `i=N` after the loop.

This rule captures both sequential and parallel loops: a parallel loop is simply one where the invariant `I` is empty. The disjoint property of permissions enforces that no data races, or other concurrency bugs due to conflicting memory accesses can occur in a parallel loop. This allows it to safely model a fork/join.

Although it provides the base of a memory safe parallel loop, some modifications, as well as a new type of group permission which we call `DesyncGroup`, are required to reflect GPU semantics. The `thread for` rule with this new permission, denoted `desync_for`, is shown in Listing 4.13.

```
t(i):
consumes H0(i), ThreadsCtx(chunk(i,N,t0..tf))
produces H(i), ThreadsCtx(chunk(i,N,t0..tf))
──────────────────────────────────────────────────────────── T-For
thread for (int i = 0; i < N; i++) { t(i) }:
consumes (desync_for i in 0..N → H0(i)), ThreadsCtx(t0..tf)
produces (desync_for i in 0..N → H(i)), ThreadsCtx(t0..tf)
```

Listing 4.13: `thread for` proof rule.

First, the "join" part of a "fork/join" means that threads automatically synchronize and communicate at the end of the loop. The resources split across threads are rejoined and can be manipulated outside the loop as though only one thread owns them all. Executing another loop spawns entirely new threads, thus the resources can be reassigned to threads in any way.

By contrast, for GPU threads, we would like synchronization to be controlled explicitly by the user, as we have seen in Section 3.2. More specifically, we would like a barrier to be necessary only when the threads change ownership of resources between loops.

This is the role of DesyncGroups, denoted `desync_for` in T-ThreadFor. DesyncGroup pins parallel loop iterations (threads) to the resources owned by them. Unlike a Group, which does not define a notion of order, it cannot be shuffled between the loops such that the ownership of resources changes. In turn, syncing entails converting DesyncGroup to Group. We discuss the implementation of barriers (which perform synchronization) in Section 4.4.

By both consuming and producing a DesyncGroup, `thread for` ensures that once some threads are given a resource, other threads cannot own that resource without using a barrier to communicate. Note that Group can always be weakened to DesyncGroup; DesyncGroup is a strictly weaker view of memory, that assumes some unspecified threads still own the resources inside.

Practically, the way that DesyncGroup prevents the shuffling/reordering of resources inside is by simply not defining the ghost operations that enable those rewriting steps on Groups. DesyncGroups and Groups are represented exactly the same in OptiGPU, as a permission with the pure type `r: range -> (items: int -> HProp) -> HProp`. `items` defines an ordered map from indices to the permissions within the group; ghosts are thus required to modify this order.

Also, although not shown here, note that `thread for` (as well as the standard parallel for) allows read-only fractions of resources to be shared between iterations. This is due to the property that having a non-full fraction of a resource means that there is no full fraction of that resource elsewhere. Thus, none of the threads can exclusively own or write to the resource that is being split by fraction across each thread.

In addition to using DesyncGroups instead of Groups, `thread for` manipulates the `ThreadsCtx` permission, which indicates what GPU threads are active. Because a `thread for` logically "forks" a set of cooperating threads, it takes a `ThreadsCtx` outside and splits it into evenly tiled chunks on the inside. This is captured by the `chunk` operation in the rule, defined as `chunk(i,N,t0..+M) = t0 + i * M/N ..+ M/N`.

This chunk corresponds directly with the size of the DesyncGroup and is how we ensure that the memory state described by the presence of DesyncGroups in a resource aligns with that of the real GPU threads.

Note that this notion of chunking allows for a flexible view on threads. Unlike other safe GPU languages [3, 13], we do not build in the concept of warps, the `x` `y` and `z` dimensions of CUDA, or even blocks. We observe that each of these are just different levels of perfectly tiled intervals in the entire grid of threads. The same is true for constructs being introduced into new GPUs, such as warpgroups (128 aligned threads). Relying solely on arithmetic i.e. divisibility of intervals means we can use multiple views on threads in the same program. For example, if it easiest to write one part of a kernel using `x` and `y` thread dimensions, but another part uses warp specialization (a technique used in high-performance kernels), we can support this with simple arithmetic rewrites on the `ThreadsCtx` permission.

## 4.3 Memory Operations

Each standard memory operation (get, set, alloc, free) is parameterized by the type of memory it works on (`mt`), to deal with the different memory types within the GPU. The list of memory types supported by OptiGPU is shown in Table 4.3.

| Name | Description |
| --- | --- |
| `Any` | default, generic memory |
| `GMem` | GPU global memory |
| `SMem` | shared memory |
| `TReg` | thread-local registers |

Table 4.3: Memory types.

The standard "points to" permission is also parameterized by a memory type; the operations for some memory type use only the corresponding permission. We use the syntax `p ⇝[mt] v` to mean the pointer `p` holds value `v` of memory type `mt`.

The `Any` memory type is the "default" and corresponds to an abstract memory that does not necessarily need to exist in hardware. We use `p ⇝ v` as syntax sugar for `p ⇝[Any] v`. Although `Any` is not strictly a "CPU memory", as a shortcut we assume that `Any` operations are legal anywhere outside GPU device code. Recall from Section 3.4 that to print a valid GPU program, all default memory operations need to be refined to use GPU memories.

We discuss the contracts for a set of highlighted memory operations on the various memory types, based on their unique restrictions stemming from their semantics on GPUs.

*Global memory.* The contracts for GPU global memory (`GMem`) are shown below.

```
Optiλ
1 template <typename T> T __gmem_get(T*
  p):
2 requires v: T, t: int;
3 reads ThreadsCtx(t..+1);
4 reads p ~>[GMem] v;
5 ensures _Res = v;
```

```
Optiλ
1 template <typename T> void __gmem_set(T*
  p, T v):
2 requires t: int;
3 preserves ThreadsCtx(t..+1);
4 writes p ~>[GMem] v;
```

```
Optiλ
1 template <typename T> T*
  gmem_malloc1(int N1):
2 preserves HostCtx;
3 produces for i in 0..N1 -> &_Res[i]
  ~>[GMem] UninitCell;
4 produces Free(_Res, ''    ''   '');
```

```
Optiλ
1 template <typename T> void gmem_free(T*
  p):
2 requires H: HProp;
3 preserves HostCtx;
4 consumes Free(p, H);
5 consumes H;
```

```
Optiλ
1 template <typename T> void
  memcpy_host_to_device1(T* dest, T* src,
  int N1):
2 requires A: int -> T;
3 preserves HostCtx;
4 reads for i in 0..N1 -> &src[i] ~>
  A(i);
5 writes for i in 0..N1 -> &dest[i]
  ~>[GMem] A(i);
```

```
Optiλ
1 template <typename T> void
  memcpy_device_to_host1(T* dest, T* src,
  int N1):
2 requires A: int -> T;
3 preserves HostCtx;
4 reads for i in 0..N1 -> &src[i]
  ~>[GMem] A(i);
5 writes for i in 0..N1 -> &dest[i] ~>
  A(i);
```

Listing 4.14: Contracts for global memory functions.

Before discussing the contracts, we note two aspects of the presentation.

First, we use C++ template syntax here as notation for a polymorphic function in our C-style Optiλ, because memory operations are polymorphic over datatypes. We are only using this syntax to express some `T` as a type argument to an Optiλ function, not to indicate C++ template semantics.

Second, certain functions like `gmem_malloc` and `memcpy_` ... end in "1". This is because Optiλ does not currently support variadic functions, and these functions should support arrays up to an arbitrary number of dimensions. In practice, our solution is to define a different version of the function up to a fixed number of arguments, that we expect programs not to exceed. For brevity, we have only shown the implementation of each of these functions for 1 dimensional arrays.

To start, `__gmem_get` (load) preserves a fraction of a permission with memtype `GMem`, and `__gmem_set` (store) writes to a `GMem` permission. Since get and set are only legal within the view of a single thread, they require a `ThreadsCtx` permission of size 1 (recall that `t0..+1` is shorthand for `t0..t0+1`).

Since global memory is allocated and copied on the host, the global memory versions of `malloc`, `free`, and `memcpy` request a `HostCtx` instead of `ThreadsCtx`. Recall that `HostCtx` is consumed by the kernel launch, meaning that these operations must occur outside device code.

The `Free` permission produced by `gmem_malloc` and consumed by `gmem_free` has the purpose of tracking the pointer that belongs to an allocated permission. On line 4, `gmem_malloc` produces `Free` with the arguments of the returned pointer, and the permission produced for that pointer (the one produced on line 3, which we show as '' '' '' to mean the same permission). Lines 4 and 5 of `gmem_free` then ensure that the permission consumed actually concerns the pointer `p` in the argument.

For memcpy, we have two versions, one for each direction between the host and device. Note that we manipulate resources of two different types; this captures the cross-memory transfer facilitated by `cudaMemcpy`.

*Shared memory.* Next, shared memory (`SMem`) follows similar restrictions as global memory for get and set (Listing 4.15); they must run on a single thread, thus `ThreadsCtx(t..+1)`, and operate only on `SMem` cells.

```
OptiÎ»
1 template <typename T> T __smem_get(T*
  p):
2 requires v: T, t: int;
3 reads ThreadsCtx(t..+1);
4 reads p ~>[SMem] v;
5 ensures _Res = v;
```

```
OptiÎ»
1 template <typename T> void __smem_set(T*
  p, T v):
2 requires t: int;
3 preserves ThreadsCtx(t..+1);
4 writes p ~>[SMem] v;
```

Listing 4.15: `get` and `set` contracts for shared memory.

However, shared memory allocation in CUDA happens in a unique fashion: The host must decide at the time of the kernel launch how many bytes are allocated for the kernel. Although only the device can access shared memory, it cannot allocate or free the memory. These semantics are captured in the rules for `__smem_malloc` and `__smem_free` (Listing 4.16)

Optiλ

```
1 template <typename T> T* __smem_malloc1(int N1):
2 requires bpg: int, sz: int;
3 preserves KernelSetupCtx;
4 reads KernelParams(bpg,_,_);
5 produces desync_for i in 0..bpg -> for i in 0..N1 ->
6   &_Res[DMINDEX1(bpg,i)][i] ~>[SMem] UninitCell;
7 produces Free(_Res, ''    ''    '');
8 consumes SMemAllowance(sz + sizeof(T)*N1);
9 produces SMemAllowance(sz);
```

Optiλ

```
1 template <typename T> void __smem_free1(T* p, int N1):
2 requires bpg: int, sz: int;
3 preserves KernelTeardownCtx;
4 reads KernelParams(bpg,_,_);
5 consumes desync_for i in 0..bpg -> for i in 0..N1 ->
6   &p[DMINDEX1(bpg,i)][i] ~>[SMem] UninitCell;
7 consumes Free(p, ''    ''    '');
8 consumes SMemAllowance(sz);
9 produces SMemAllowance(sz + sizeof(T)*N1);
```

Listing 4.16: Contracts for shared memory `malloc` and `free`.

The rules implement these semantics with 3 mechanisms.

First, `__smem_malloc` can only be executed in a `KernelSetupCtx`, which we recall is the execution context available in between host code and the start of the device code (where threads can be spawned), called the setup phase. Conversely, `__smem_free` can only be executed in the teardown phase, before the `kernel_kill`, but after GPU code. These requirements mean allocating or freeing shared memory happens only with the kernel launch.

Second, the number of bytes allocated is checked using the SMemAllowance permission, for which `__smem_malloc` decreases by the size of the memory allocated, and `__smem_free` increases. Because the launch produces SMEMAllowance with the number of shared memory bytes provided as argument, this ensures only the requested amount of shared memory can be allocated. Note that this is the reason we to pass the dimensions to `__smem_free`, unlike `__gmem_free`; the pointer in the `Free` permission does not give us any information about the allocated size. We have opted to add this information to the arguments of `__smem_free` instead of the `Free` permission.

Third, `__smem_malloc` produces a DesyncGroup of the memory requested with the number of blocks specified in the kernel. This means that each block has its own chunk of memory of the size requested. Since it is already in a DesyncGroup, it also means that each block's shared memory cannot be reassigned to another block.

In a way, `__smem_malloc` treats the shared memory as one contiguous array for the entire grid, but due to the DesyncGroups, each block has its own equally sized chunk of the array that only it can access. However, we need to track the extra dimensions for indexing the block's chunk within the contiguous array, because those dimensions should be erased to leave only the block's local view after CUDA code generation.

We use `DMINDEX` for this; `DMINDEX` syntactically differentiates the "distributed" part of an index into an array with the regular part. `__smem_malloc` produces a permission with a standard multi-dimensional array access, but with a `DMINDEX` in the outermost dimension, corresponding to the DesyncGroup over the blocks. `DMINDEX` is used when reasoning about correct accesses into the contiguous array (since it is expressed in the permission), but erased (by replacing it with 0) at code generation (explained further in Chapter 6). Also, `DMINDEX` is another variadic function for which we have simply defined multiple versions, thus we use `DMINDEX1` for the 1 dimension in the contract.

*Thread registers.* Finally, we discuss thread-local registers (`TReg`). For this, we only show the rule for `__treg_malloc` (Listing 4.17), as `get` and `set` are once again the same, and `free` happens automatically (more below).

OptiΛ

```
1 template <typename T> T* __treg_malloc0():
2 requires t: int, sz: int;
3 preserves ThreadsCtx(t..+sz);
4 produces desync_for i in 0..sz → &_Res[DMINDEX1(sz,i)] ⤳[TReg]
  UninitCell;
```

Listing 4.17: Contract for thread register memory allocation. Once again, allocation functions are variadic, and we only show one version, in this case for a 0-dimensional array (scalar local variable).

Similarly to shared memory, when allocated, thread registers are wrapped in a DesyncGroup corresponding to the size of the current thread context. Without this mechanism, i.e. if thread local registers could only be declared inside the scope of a thread level loop, they could not be preserved across across collective operations outside loops, like barriers. Thus, in these situations, thread registers are allocated at the collective level instead. Listing 4.18 shows an example of this pattern in CUDA, and the corresponding OptiGPU program, which allocates the variable `x` in the CUDA program as an array corresponding to the current `ThreadsCtx` size.

```c
__global__ void kernel(int *A) {
  int t = threadIdx.x;
  int b = blockIdx.x;
  int x = A[b * blockDim.x + t];
  __syncthreads();
  int t1 = (blockDim.x - t - 1);
  A[b * blockDim.x + t1] += x;
}
```

```c
thread for (int b = 0; b < bpg; b++) {
  int* const xs = __treg_malloc0<int>();
  thread for (int t = 0; t < tpb; t++) {
    xs[t] = A[b * tpb + t];
  }
  blocksync();
  thread for (int t = 0; t < tpb; t++) {
    A[b * tpb + (tpb - t - 1)] = xs[t];
  }
}
```

Listing 4.18: Example of a local register hoisted out of the thread loop and allocated as an array instead. The barrier is necessary because the second loop accesses the array backwards.

Collective allocation of thread registers uses the same mechanism as allocating shared memory at the grid level, and thus the extra dimensions are also automatically erased at code generation.

As mentioned, `TReg` does not provide a `free` operation, because it is a form of stack allocation. OptiTrust's typechecker consumes the permissions corresponding to stack allocations at the end of a scope automatically. Currently, this is implemented by a syntactic check for allocations when typechecking a scope. We have extended the typechecker to check for `__treg_malloc` as well.

## 4.4 Barriers

We show the specification for a block level barrier (`__syncthreads` in CUDA) in Listing 4.19.

OptiΛ

```
1 void blocksync():
2 requires H: HProp, tpb: int, t: int;
3 preserves ThreadsCtx(t ..+ tpb);
4 reads KernelParams(_, tpb, _);
5 consumes H;
6 produces Sync(block_sync_mem, H);
```

Listing 4.19: Contract for a block-level barrier.

To enforce that the barrier only happens at a block level, it takes a `ThreadsCtx` (line 3) with the size equal to the number of threads per block `tpb`, specified in the `KernelParams` permission (created at launch). Since the `KernelParams` permission is shared over all the blocks in the grid, the barrier takes only a read-only fraction (line 4).

Recall that with collective operations being executed above the thread level, it is not syntactically possible to introduce thread divergence. The `ThreadsCtx` requirement enforces that the collective operations are executed by the expected number of threads.

On line 5, we consume `H`, an arbitrary permission we have quantified over on line 2. This is the permission that we would like to transform to reflect a synchronized memory state. Barriers accomplish this by converting DesyncGroup to Group, allowing resources to be reassigned to threads between `thread fors`.

To facilitate this conversion, we use the permission `Sync(fM, H)`. `Sync` is a meta-level operation that converts DesyncGroups to Groups in-depth; we encode it as a permission itself to avoid needing to reflect on resources in the contract for barriers, something that Opti$\lambda$ does not support. Instead, we define the operation of `Sync` as another set of ghosts/reasoning rules:

$$\begin{aligned} \texttt{Sync(fM, p ~>[M] v)} &= \texttt{p ~>[M] v} \text{ if } \texttt{fM(M)} \\ \texttt{Sync(fM, for i in 0..N -> H)} &= \texttt{Sync(fM, for i in 0..N -> H)} \\ \texttt{Sync(fM, desync_for i in 0..N -> H)} &= \texttt{Sync(fM, for i in 0..N -> H)} \end{aligned} \tag{4.1}$$

One issue is that certain barriers should only be able to re-synchronize certain memory types. In particular, it should never be possible to re-synchronize thread registers: these are always tied to their respective threads. This is the role of `fM: MemType -> Prop`. Note that `Prop` is the type of propositions – one of the types of pure resources – in Opti$\lambda$. The memory type of the inside resource must satisfy the predicate `fM`. This mechanism can also be used to describe barriers in other languages like OpenCL, which provide barriers that e.g. work only on shared memory.

To avoid noise in proof scripts, the OptiGPU typechecker applies the simplifcation rules above eagerly on `Sync` resources in the context. If `Sync()` is used incorrectly (on the wrong memory type), it will get stuck at `p ⇝[mt] v`.

Going back to the example of `blocksync()`, it produces `Sync(block_sync_mem,H)`. We do not show its definition here, but `block_sync_mem` is a predicate for which the axioms `block_sync_mem  GMem` (read as "`GMem` satisfies `block_sync_mem`") and `block_sync_mem SMem` hold. In other words, `blocksync()` only synchronizes global and shared memory.

We also define barriers at the end of the kernel. For synchronous CUDA kernel launches, there is an implicit barrier of all global and shared memory at the end of the kernel. Thus, we define a `kernel_teardown_sync()` which happens at the teardown phase (Listing 4.20).

OptiΛ

```
1 void kernel_teardown_sync():
2 requires H: HProp;
3 reads KernelTeardownCtx;
4 consumes H;
5 produces Sync(block_sync_mem, H);
```

Listing 4.20: Contract for barrier in kernel teardown phase.

The contract for `kernel_teardown_sync` is mostly the same as `blocksync`, only that the expected execution context is different.

Finally, we note that the contracts for barriers only take one permission; in reality, a CUDA barrier synchronizes all memory of the given type in one call. Our implementation is due to current limitations in OptiΛ in being able to quantify over multiple resources in a contract. Translating each OptiGPU barrier call to a CUDA barrier call is likely not unsound, because the CUDA barriers would only synchronize *more* memory, but it is undesirable for performance – the OptiGPU barrier calls should all be collapsed into one. As a workaround, we simply collapse adjacent calls to the same barrier function during code generation. This is described further in Chapter 6.

# 5 Transformations

Recall that in OptiGPU, we aim to maximize reuse of OptiTrust's existing transformations by refining a serial, imperative program as much as possible first, and then converting it, in particular parallelizing it to run on a GPU. OptiGPU only adds a set of transformations that perform this "GPU conversion" process. These transformations are the topic of this section.

A central challenge with this conversion process is handling the gap between the semantics of the final GPU program and the source CPU program. A correct GPU program in OptiGPU has a number of *non-local* dependencies. For example, using GPU memory operations requires establishing `thread for`s to expose a `ThreadsCtx`. But to use `thread for`, there must be a `ThreadsCtx` to begin with, which means that a kernel launch must be inserted somewhere surrounding the loop. In addition, since `ThreadsCtx` is dependent on the `thread for` loops surrounding a given term, partially converting sequential loops to `thread for` loops could break the proof if done in the wrong order.

Whenever possible, OptiTrust prefers the use of localized, peephole transformations, from which complex transformations can be composed. Local transformations give more feedback for the user - finer grained errors on what step a proof went wrong - and are easier to maintain and reuse. Yet, the non-local dependencies in an OptiGPU program make local transformations a comparatively non trivial implementation task for our needs. For example, whereas `thread for` conversion requires reasoning about the surrounding loops, the existing "sequential for to (CPU) parallel for" transformation in OptiTrust (seen in the reduce example of Section 2.2.1) simply changes the mode of the loop; if the loop is parallelizable, no extra work is required for the proof to be preserved.

We employ a variety of strategies in OptiGPU to cast GPU conversion as a series of local, and correct source-to-source rewrites wherever possible. We discuss the other strategies throughout this section, but the most relevant is that we expose these smaller transformations while designing them to be grouped in *phases* of transformations. Within a phase, the transformations do not guarantee correctness, and typechecking should be temporarily suspended. However, the end of each phase represents a clear step in the GPU conversion process, which we will be able to typecheck and provide feedback to the user on, even if the program is not syntactically valid according to the CUDA code generator. For example, parallelizing sequential loops as `thread for` is its own phase; by the end of this phase, it is expected that the thread hierarchy has been described and is correct with respect to the kernel launch. Within the phase, it is possible that the program is not valid, due to only partially converting `thread fors`, for instance. Memory conversion follows this phase, as GPU memory operations require `ThreadsCtx` to exist.

Figure 5.3 depicts the phases of GPU transformations. We now discuss each phase and the transformations within in more detail.

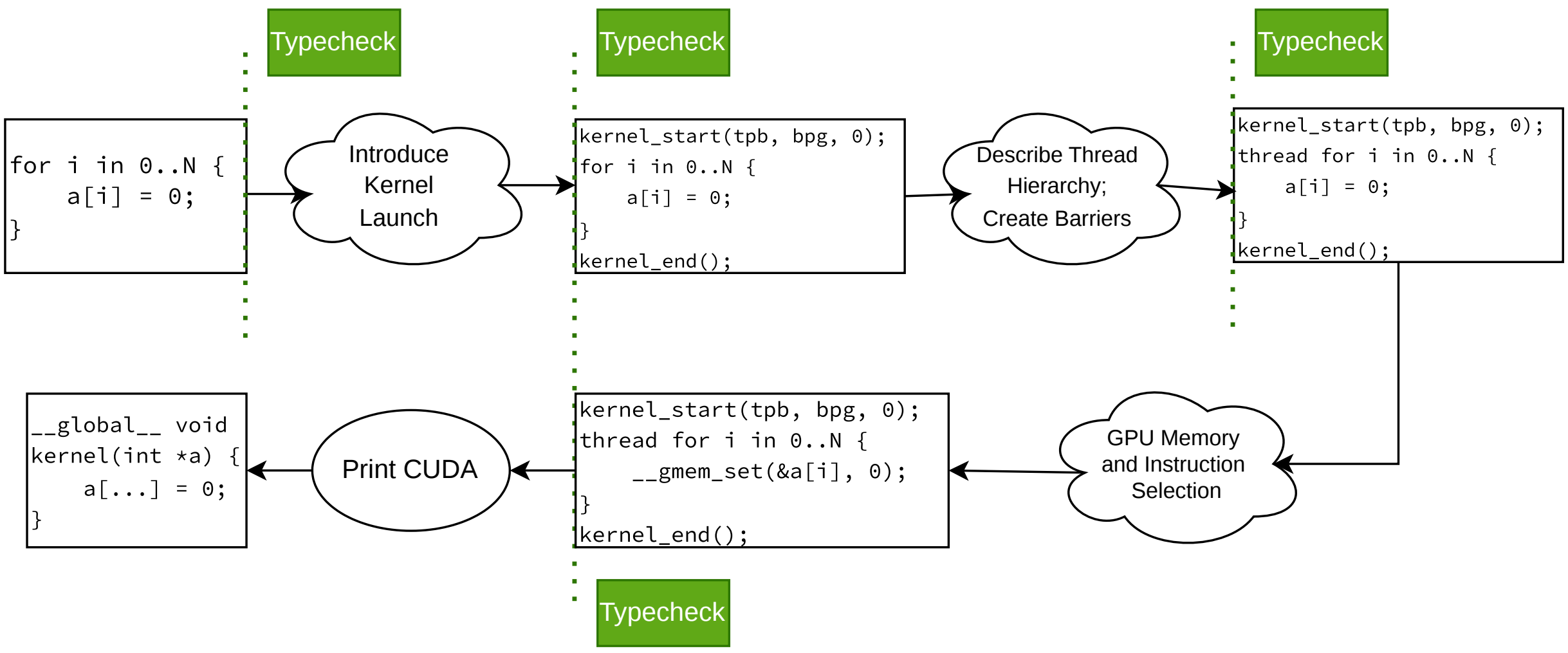

Figure 5.3: Workflow of GPU transformations with phases in OptiGPU. Phases are represented by the bubbles in between code boxes. The green dashed lines show points in between phases, where the program is expected to typecheck.

## 5.1 Introducing Kernel Launches

Note that before any of these phases, we expect all of the program's structure and optimizations to be derived, only in the form of a serial CPU program. This is represented by the leftmost box in Figure 5.3, although of course it is a trivial example that does not show CPU optimizations.

After that, the first phase is to add a kernel launch, which is shown in the middle box of Figure 5.3. This ordering is based off the observation that kernel launches do not themselves have resource dependencies on anything else, whereas the other constructs depend on the resources produced by a kernel launch.

The transformation `Gpu.create_kernel_launch` takes a target to any term in the AST, as well as terms for the arguments to `kernel_launch`, and wraps the term in a sequence (i.e. a scope with curly braces `{ }`) containing the transition functions at the beginning and end.

This transformation won't break the proof of the program at all, it only introduces the execution contexts required to run GPU code, which are unused. The only caveat is that if the function does not require the `HostCtx` execution context as a precondition, launching a kernel would of course fail because the fact that the function is running on the host has not been made explicit. Because it concerns modifying the assumptions of the function, OptiGPU expects the user to specify this manually.

## 5.2 Describing Thread Hierarchy

The second step is to introduce the structure of the thread hierarchy, by converting the standard for loops into `thread for` loops. This is shown in the top right box of Figure 5.3.

As alluded to at the beginning of this chapter, `thread for` conversion is not as straightforward as CPU parallel for conversion due to the `ThreadsCtx` requirements; it is not correct to convert a sequential for in a nest in any order. . In addition, however, recall that `thread for` modifies resources after the loop to be in a "desynchronized" state (with `DesyncGroup`). Without proper care to insert barriers, ghosts in between adjacent loops that facilitate the reassignment of threads to memory resources will fail due to expecting `Groups` instead of `DesyncGroups`.

One strategy OptiGPU employs here is to split these concerns by introducing a `magic thread for`. We omit its full contract here for brevity, but it is identical to a `thread for` minus the `ThreadsCtx` requirement. The transformations start by converting sequential for to `magic thread for`, which first establishes the notion of barriers and `DesyncGroups` in the program, and then converts these loops to `thread for` all at once. `magic thread for` is not recognized by the CUDA code generator and only serves as an intermediate reasoning step.

`magic thread for` allows us to cast thread for conversion as a peephole transformation, the same way CPU parallel loop conversion is: the transformation always inserts a barrier directly after the loop, thereby resynchronizing resources immediately. For this, we also need a `magic_barrier`: one that creates a `Sync(h,fM)` with an `fM` that is true for all memory types, and that does not require a `ThreadsCtx`. `magic_barrier` is also not recognized by the code generator, and must be converted to a realizable GPU barrier beforehand.

Of course, adding a barrier after every `thread for` defeats the purpose of explicit barriers. At some point, the transformations need to deal with modifying contracts in the entire nest of loops to reflect that `DesyncGroups` are produced . However, *starting* with a barrier allows us to reason about moving it to where it is actually needed in the loop nest, or possibly removing it eventually, in small, correctness-preserving steps. This avoids having to implement a "monolithic" transformation that needs to transform the entire loop nest at once without typechecking in between steps.

We now elaborate on these small steps that facilitate `magic thread for` conversion, and how they are put together in a user-facing combined transformation. The implementation is depicted in Figure 5.4.

The combined transformation works at the granularity of *tail loop nests*, which are loop nests where each loop except the outermost is only proceeded sequentially by ghosts or the end of the enclosing loop body. This is based on the observation that loops in "tail position" like this should not need a barrier, and any work they are doing in the form of ghosts that manipulate groups can be postponed until after a barrier in an enclosing scope, where the barrier is more likely to actually be necessary. For instance, the initial program in Figure 5.4 a) (at the top left) has a tail nest of loops over `i` and `j`, because the `j` loop is directly proceeded by the end of the `i` loop. Note that converting tail nests is only a heuristic that is designed to minimize trivially unnecessary barriers. As we have mentioned, the user can always manually insert their own proofs and transformations to optimize barrier placement if necessary.

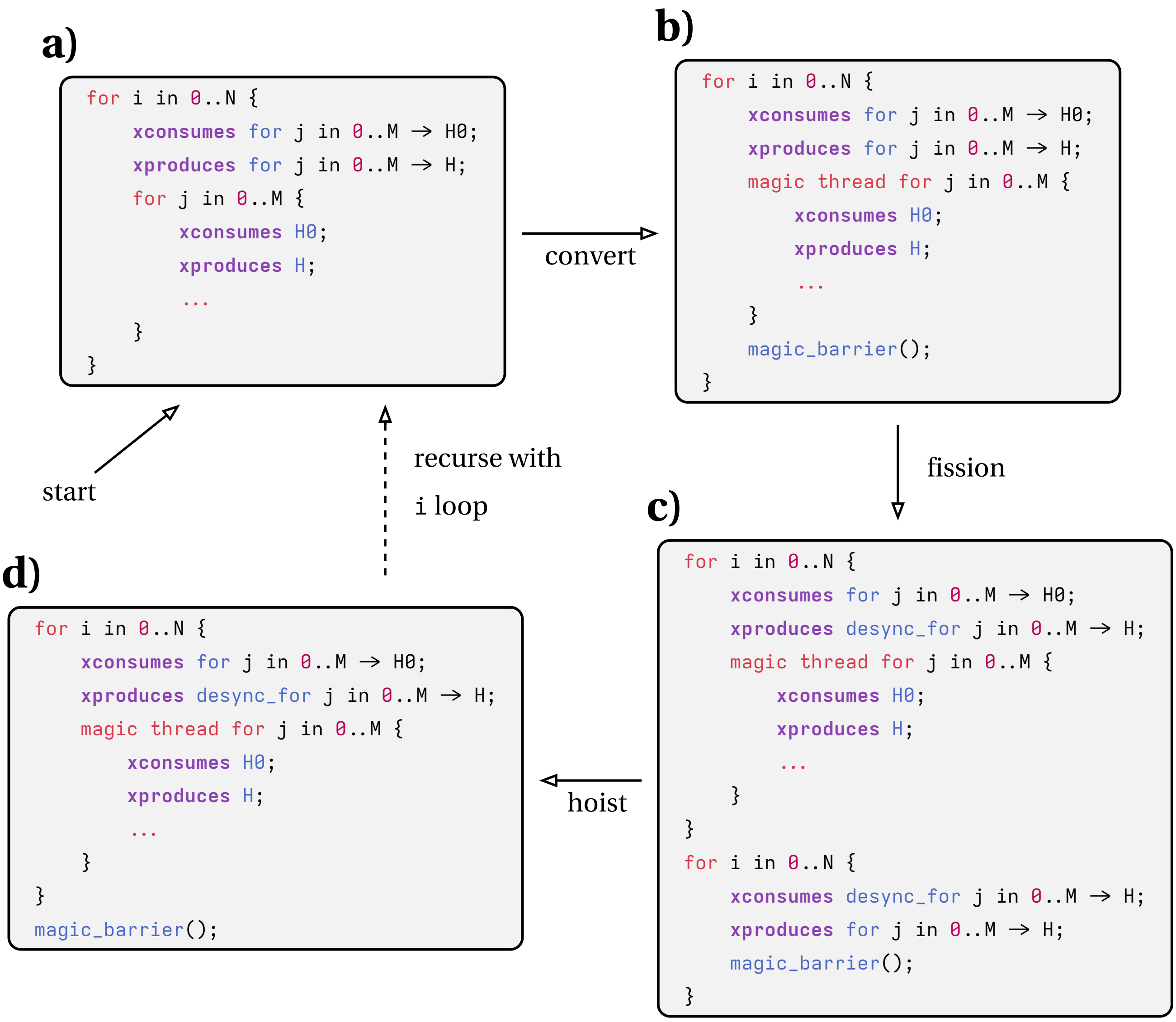

Figure 5.4: Small transformation steps of `magic thread for` conversion.

Loops in the nest are converted recursively, starting with a target to the innermost loop; in the case of . The innermost loop is always converted, but the user passes a list of booleans that indicate, for the remaining loops in the nest, whether or not that loop should remain sequential or become `thread for`. If a loop is to be converted, the transformation changes the mode to `magic thread for`, and adds a `magic_barrier` directly after, unless there already is one. Conversion is implemented by the basic transformation `seq_for_to_magicthread_for`, abbreviated "convert" in the arrow between a) and b) in Figure 5.4).

Then, the combined transformation attempts to "hoist" this barrier out of the targeted loop and at the tail of the enclosing one. This is done by first applying the standard `Loop.fission` transformation in OptiTrust at the point in between the end of the targeted loop and the barrier directly after, and then again between the barrier and the rest of the loop body (if there are any instructions). The arrow labeled "fission" between Figure 5.4 b) and c) demonstrates this transformation. With no awareness of GPU operations, `Loop.fission` takes care of modifying the contracts to reflect the `DesyncGroups` now produced by the converted loop. This is why breaking into small, correctness-preserving steps is useful: because `Loop.fission` requires typing information, we otherwise would have had to implement `Loop.fission` ourselves with the added constraint of not having typing information.

Now, the barrier is in its own loop. But a loop with a barrier inside is equivalent to a barrier itself – recall that `Sync` converts `DesyncGroups` in depth. The basic transformation `remove_loop_around_barrier` applies this rewrite, shown in the arrow labeled "hoist" between Figure 5.4 c) and d).

At this point, we recurse with the innermost loop set as the enclosing loop of the previous one. This is depicted with the final arrow of Figure 5.4 going from d) to a). The recursion proceeds until the list of loops to convert is empty, meaning we have reached the outermost loop. At this point, we have collected any ghosts that appeared in these loops and moved them after the barrier, which is now at the outermost point.

For now, the user manually breaks their program into tail loop nests and calls this combined transformation on each. In the future, we plan to further combine transformations so that the user only provides a target to the body of their kernel, and a target that matches any loop which should *not* be converted, and the rest is handled automatically.

After converting all necessary loops to `magic thread for`, the last step for the user is to apply `convert_magic_thread_fors`, which takes a target to the outermost loop in the kernel, and transforms every `magic thread for` inside to `thread for`. If the hierarchy has been set up correctly to match the `ThreadsCtx` provided by the kernel, no other work is necessary.

On the other hand, converting `magic_barrier` is specific to the operations available on the target GPU (a warp-level sync, a block sync, or the sync at the end of the kernel), and is thus left for the final phase. It is also easier to convert `magic_barrier` after converting memories; real barriers do not accept the default `Any` memories that are still present at this point.

## 5.3 GPU Memory and Instruction Selection

The last step in the process, shown in the bottom center box of Figure 5.3, is to realize the memory operations and barriers in the program as real operations on the GPU.

The memory transformations take a target to a declaration, and change all operations, such as get, set, alloc, free, etc. to use the appropriate versions for that memory type. They must also modify the resources accordingly, so that `p ⤳ v` becomes e.g. `p ⤳[SMem] v`.

In the case of shared memory and thread registers[4], as explained in Section 4.3, certain dimensions are distributed dimensions, in that they correspond to thread indices, instead of being realized in the memory on the hardware. In these cases, the transformations erase these dimensions in the allocation, with the number of dimensions to erase being provided by the user. They also modify all accesses in that memory to use `DMINDEX`. If the user has set up the allocation properly and erased the correct number of dimensions, these erased dimensions should match what the current threads context is, and the program will still typecheck.

Finally, the magic barriers from before are converted to real constructs. The user specifies a target to the magic barrier and the type of barrier to replace with (block, warp, kernel end, etc.). If the barriers are in legal positions with respect to `ThreadsCtx`, no extra justification is required.

[4] We note that while thread registers are supported in the language, we have not yet implemented the conversion transformations for them.

At this point, only constructs that can be generated to CUDA should remain, which is depicted in the bottom left area of Figure 5.3. This code generation is the subject of the next chapter.

# 6 CUDA Code Generation

OptiGPU implements a CUDA code generation pass that is executed at the end of the transformation script. It is implemented as a source-to-source transformation on the internal AST (Optiλ), although it is *not* verified, as OptiGPU cannot typecheck CUDA itself. The rationale for this implementation is that OptiTrust can already generate most of what is needed for CUDA in its existing Optiλ to C++ printer. Thus, the CUDA code generator turns the native GPU constructs of Optiλ to "CUDA-like" Optiλ code, and passes that AST to the existing C++ printer. For CUDA-specific C++ extensions like the `__global__` keyword, style annotations are added to the AST nodes. We briefly describe the source-to-source side of the code generation process in this chapter.

First, note that during printing, all ghosts, specifications, and contracts are removed, and only executable code remains. The transformation to the "CUDA-like" AST then proceeds in the following steps.

*Identifying device code.* A sequence of terms beginning with `kernel_start`, with `kernel_setup` and `kernel_teardown` inside, and ending with `kernel_end` denotes a kernel launch. The kernel body, i.e. the sequence between setup and teardown, must be lowered according to the steps that follow. Once it is lowered, however, the free variables of the lowered body with respect to the enclosing function (performing the call) are computed. Then, it is lifted into a separate function, which is annotated as a kernel (with `__global__`) having arguments as the free variables calculated.

Correspondingly, the original sequence is replaced with a launch (with the triple chevron) to the new kernel, with each argument bound to the instantiation of the free variables outside the sequence. The parameters given in the triple chevrons also correspond exactly to those passed in `kernel_launch`: number of blocks per grid, threads per block, and number of bytes of shared memory. Note that we always use `1` dimension for blocks and threads when generating CUDA; we compute indices for other dimensions inside the kernel from `threadIdx.x` and `blockIdx.x`.

Next, any calls to `smem_malloc` that appear in the setup phase are added as declarations in the the beginning of the kernel. CUDA supports both dynamic shared memory allocation (from the host) and static allocation. The static form is expressed as a standard C local variable declaration with the `__shared__` attribute, e.g. `__shared__ int A[10]`. For simplicity, and since our case studies so far have not utilized dynamic allocation, OptiGPU currently generates all `smem_mallocs` in this form, even though its arguments may be dynamic. Supporting dynamic allocation is a matter of using one `extern __shared__` declaration and setting each shared memory pointer as offsets into this single buffer for the whole shared memory. We plan to implement this dynamic allocation support in OptiGPU in the future.

Finally, any other code that appears between `launch` and `kill` but not in the kernel body is erased and considered not executable. Most of this code consists of ghosts for rewriting resources, but we do mention `kernel_end_sync` and `smem_free`, which both appear in the teardown phase. These are both legal to erase because they are handled implicitly by the CUDA runtime when doing a launch: a kernel implicitly synchronizes when it finishes, and shared memory is implicitly deallocated at the end of a kernel.

*Projecting `thread for` onto single-thread code.* Recall that the principle of `thread for` is that even in collective contexts (outside the inner most loop), it should be correct to project any code inside a loop to a single-thread view.

Thus, inside the kernel body extracted by the previous step, every `thread for` is replaced with its body. Within the body, the index variable of the loop is substituted with a new variable, declared at the start of the kernel, which calculates the value of the loop index from the global thread index provided by CUDA. By global index, we mean within the entire grid, i.e. the result of `blockIdx.x * blockDim.x + threadIdx.x` using CUDA variables. The index for a given loop is the global index divided by the size of the `ThreadsCtx` for that loop, but since typing information is not available, it is calculated by dividing the grid size by the bounds of all enclosing loops.

Listing 6.21 shows an example of CUDA code (device side only) generated by OptiGPU, from our transpose case study, showing the indexing expressions that we generate to handle `thread for`.

```cpp
1 __global__ void __kernel0 (float* d_b, float* d_a, int H, int W)  {
2   const int __ctx_sz = MSIZE2(exact_div(H, 32), exact_div(W, 32)) * MSIZE2(16, 32);
3   const int __tid = blockIdx.x * MSIZE2(16, 32) + threadIdx.x;
4   __shared__ float tile[MSIZE2(32, 32)];
5   const int __ctx_sz_0 = __ctx_sz / (exact_div(H, 32));
6   const int __by0 = __tid % __ctx_sz / __ctx_sz_0;
7   const int __ctx_sz_1 = __ctx_sz_0 / (exact_div(W, 32));
8   const int __bx1 = __tid % __ctx_sz_0 / __ctx_sz_1;
9   const int __ctx_sz_2 = __ctx_sz_1 / 16;
10   const int __y2 = __tid % __ctx_sz_1 / __ctx_sz_2;
11   const int __ctx_sz_3 = __ctx_sz_2 / 32;
12   const int __x3 = __tid % __ctx_sz_2 / __ctx_sz_3;
13   const int __ctx_sz_4 = __ctx_sz_1 / 16;
14   const int __x4 = __tid % __ctx_sz_1 / __ctx_sz_4;
15   const int __ctx_sz_5 = __ctx_sz_4 / 32;
16   const int __y5 = __tid % __ctx_sz_4 / __ctx_sz_5;
17    for (int j = 0; j < 2; j++) {
18     tile[MINDEX3(exact_div(H, 32) * (exact_div(W, 32)), 32, 32, 0, j * 16 + __y2, __x3)]
  = d_a[MINDEX2(H, W, __by0 * 32 + (
19       j * 16 + __y2), __bx1 * 32 + __x3)];
20   }
21   __syncthreads();
22    for (int j = 0; j < 2; j++) {
23     d_b[MINDEX2(W, H, __bx1 * 32 + (j * 16 + __x4), __by0 * 32 + __y5)] =
  tile[MINDEX3(exact_div(H, 32) * (
24       exact_div(W, 32)), 32, 32, 0, __y5, j * 16 + __x4)];
25   }
26 }
```

Listing 6.21: Transpose kernel after CUDA code generation.

Currently, OptiGPU performs no optimization on these expressions; even simple common subexpression elimination is not yet implemented, leading to cases like `__y2` and `__x4` computing the same result in Listing 6.21. Because these computations occur at the top of the kernel – outside any hot path – their performance impact is generally minimized. However, in programs without long running loops on single threads, like transpose, the difference is more apparent, as we see in Chapter 7. Thus, applying optimizations like constant folding and common subexpression elimination in our code generation is a next step for future work on OptiGPU.

*Printing GPU operations.* There are a number of practical code generation/instruction selection tasks that we mention here.

First, while operations for the various memory types use intrinsic functions like `__smem_get` in OptiGPU, they are all the same C load/store operations in CUDA. Thus, we replace these function calls with the standard `Any` operations so they are printed correctly.

Other OptiGPU functions are translated to CUDA built-ins and library functions instead, such as cudaMemcpy/cudaMalloc for global memory, `__syncthreads` for block-level barriers, etc.

However, as mentioned in Section 4.3, barriers are actually sequences of barriers in OptiGPU, because the contract for a barrier can only convert one memory resource at a time. Generating each of these barriers as one call to e.g. `__syncthreads` would be undesirable, so when transformations generate a sequence of barriers, the sequence is tagged such that the code generator recognizes that the sequence should be collapsed into one. It merely checks that the sequence only contains function calls to one type of barrier function, and then replaces the sequence with one call accordingly.

*Erasing distributed dimensions* For thread registers and shared memory, a variable's representation within OptiGPU has more dimensions than it will have in CUDA. These dimensions are already implicit in the allocation, so the printer does not have to change there. The array accesses to the variables use the extra dimensions, but since they are separated syntactically with `DMINDEX` (Section 4.3), it is sufficient to replace all calls to `DMINDEX` with 0. Although we do not implement it on the OptiGPU side, instead relying on e.g. `nvcc`, standard constant propagation can simplify the expressions left after substituting `DMINDEX`.

# 7 Evaluation

In this chapter, we evaluate OptiGPU on the following criteria:

1. The ability to express optimizations and CUDA features utilized in real-world handwritten CUDA code
2. The performance of OptiGPU generated code versus a handwritten counterpart
3. The user effort required to write verified programs in OptiGPU, in terms of # of lines of input

To this end, we have applied OptiGPU on 2 standard CUDA kernels, serving as our case studies: matrix transposition, and parallel reduction. We outline our experimental setup (Section 7.1), then compare performance of OptiGPU and handwritten code for each of these case studies (Section 7.2), and conclude with a discussion on user effort centered on lines of input code required (Section 7.3).

## 7.1 Experimental Setup

We sourced reference kernels from the `cuda-samples` repository from Nvidia, with commit `4f73561` [17]. The experiments were ran on a system with an RTX 5060 GPU, an AMD Ryzen 5 3600X CPU with 32GB RAM, running Ubuntu 22.04 through WSL on Windows 11. We used CUDA 13.0 with the Nvidia driver 581.29, and `nvcc` version 13.0.88. All benchmarks were compiled with `nvcc` using `-O3`. Our implementation of OptiGPU is available on GitHub here .

We use *effective bandwidth* to measure performance of the kernels, calculated as $\frac{R_B+W_B}{t}$, where $R_B$ and $W_B$ are the numbers of input and output bytes processed by the kernel respectively, and $t$ is the runtime of the kernel. This follows performance measurement conventions used by Nvidia [11]. We averaged out the runtimes of kernels over 1000 runs. Although OptiGPU generates host and device code, the host-side CUDA operations like `cudaMalloc` and `cudaMemcpy` would dominate the runtime and create too much noise, so we only use the runtime of the device portion. Only the reduce benchmark performs operations other than memory transfers outside the kernel, and its host side code is identical to that of the reference kernel, so we do not believe there to be significant differences in this regard.

For the transpose kernel, we used a 4096x4096 matrix (16M elements) as input and output. The launch parameters (# of blocks, # of threads) were fixed by the size and the optimizations applied in the kernel. We compare our kernel against `transposeNoBankConflicts` in `transpose.cu`; the other kernels either perform worse or only compute a partial transpose.

For reduction, we follow the default parameters in the `reduction.cpp` test file in the `cuda-samples` repo, which a 16M element input array and 256 threads per block. We compare against two kernels in `reduction_kernel.cu`: `reduce3` is the code we tried to recreate in OptiGPU, while `reduce6` is the best performing on our system, but it contains features like warp shuffles we do not yet support.

## 7.2 Results

| Transpose | | Reduction | | |
|---|---|---|---|---|
| OptiGPU | Reference | OptiGPU | Reference | Optimal |
| 297.7 GB/s | 340.6 GB/s | 297.9 GB/s | 349.4 GB/s | 376.8 GB/s |

Table 7.4: Effective bandwidth in GB/s for each kernel tested. “Optimal” in the case of reduction represents the best performing kernel from cuda-samples (`reduce6`), while “Reference” is the kernel we tried to recreate in OptiGPU (`reduce3`).

Table 7.4 shows the effective bandwidth of the kernels for each case study. The transpose kernel and its transformation script is mostly the same as the code shown Chapter 3, although we apply an additional loop tiling step so that each thread processes 2 elements of the matrix sequentially, following the reference version. We were able to reproduce the structure of the reference kernel in OptiGPU, but our implementation performs about ~40

GB/s worse, which we measured through manual editing is due to our unoptimized generation of indexing optimizations (mentioned in Chapter 6). We plan to implement simple optimizations like common subexpression elimination and constant folding in our code generation to address these shortcomings.

The reduction kernel uses a tree-based parallel reduction, where a block-local shared memory is iteratively split in half and each thread sums an element from the first half with the second in parallel. In this way, the sum of an array of length $N$ can be computed in $\log(N)$ steps. We were able to reproduce this optimization in our OptiGPU reduction script, verifying it in the process. The only aspects of the reference kernel we were targeting that we were not able to reproduce are: 1) using a grid-stride to sum 2 elements from global memory directly on each thread, instead of summing 2 adjacent elements (what our script did instead), 2) using a temporary register as an accumulator explicitly in the tree reduction phase instead of modifying shared memory directly. Through manual editing and testing of the OptiGPU output we determined that the second aspect is the cause for the 50 GB/s gap between the reference. Specifically, the difference between the OptiGPU and reference code causing the issue is the use of `mySum` in the lines of code highlighted in Listing 7.22.

```
// Reference
...
sdata[tid] = mySum = mySum + sdata[tid + s];
...
// OptiGPU
...
reduce_arr_1[__t2] = reduce_arr_1[__t2] + (&reduce_arr_1[ei])[__t2];
...
```

Listing 7.22: Discrepancy between reference and OptiGPU generated CUDA. `mySum` is an accumulator (in thread register memory) declared outside the reduction loop.

This code is expressible in OptiGPU's language, as it supports thread registers, but we have not yet implemented support for them in transformations. We aim to address this in future work.

## 7.3 Size of Input Code

Recall that a goal of OptiGPU is to reduce proof burden. Our primary reference point for this goal is Kuiper [15], but as both Kuiper and OptiGPU are work-in-progress prototypes, we do not attempt to directly compare the proof burden required in each tool. This section attempts to give an approximate measure of OptiGPU's reduction in proof burden as a result of proof-preserving transformations, by comparing it against itself *without* using transformations. We compare lines of code (proof and non-proof) for the input, which is both the input program and the transformation script, with the output lines after transformations, which represent what a user might have to write without using transformations.

| Kernel | OptiGPU Input | | | OptiGPU Output | | | Reference CUDA (kernel only) |
|---|---|---|---|---|---|---|---|
| | C Code | Annot. | Transfo Script (OCaml) | C Code | Annot. | CUDA/ Kernel Only | |
| Transpose | 9 | 9 | 57 | 89 | 359 | 37/26 | 18 |
| Reduction | 8+24 | 14+71 | 143 | 93 | 180 | 54/37 | 20 |

Table 7.5: Breakdown of input and output code size in OptiGPU, and a comparison to reference CUDA code, for each case study.

Table 7.5 shows the breakdown of input and output lines for our case studies. Lines of code are always measured ignoring whitespace and comments. For the reference code, we only measure the number of lines of code for the kernel itself, as the host code for these programs has the kernel launch and memory transfer (what OptiGPU generates) mixed in with other logic for measuring performance, parsing user input for the kernel parameters, error handling for CUDA runtime functions, etc.

For OptiGPU, we measure 3 input variables: 1) the number of lines of non-proof or annotations in the input program ("C code"), 2) the number of proof/annotation lines ("Annot."), 3) the number of lines of OCaml code in the transformation script ("Transfo Script"). We consider any line in the C code that is part of an annotation (__) to be a proof/ annotation line. For instance, if there is a contract whose input string continues onto the next line, we consider that line a proof line as well.

For the OptiGPU output, we measure both the CUDA code generated, and the direct C printing of the Opti$\lambda$ source at the last step of the script, which resembles the syntax with e.g. `thread for` that we have seen throughout this thesis. We consider this code so we can measure the proof requirements for the final, fully transformed program. Therefore, we have the same 2 columns of proof lines and non-proof lines as the input, and a column for the generated CUDA. We also report the number of those lines pertaining only to the kernel (device-side), to allow comparison with the reference.

For reduction, our strategy to derive the tree-based parallel reduction involved writing and verifying a simple implementation of this algorithm separately, and then inserting and inlining the routine into the code with the transformation script. We found this to be easier than trying to break it down into transformations on a naive sequential reduction. As a result, for the input of the reduction kernel, we report the lines of code (both code and annotations) for the initial program and this tree reduction program separately using `+`. They are both used as part of the same output in the end, so the output column only has one number to report.

Overall, the results shown by the table are mixed, but we can see that in the case of transpose, the lines of transformations combined with the initial annotations (66) were significantly less than the annotations required in the final program (359), showing some potential for OptiGPU to automate proof obligations. As it involves a number of loop tiling, reordering and fissioning steps, the transpose kernel resulted in many more proof lines per non-proof line than the reduction kernel. These loop transformations tend to require many ghosts, and the reduction kernel used less of these. The main issue with reduction is the length of the transformation script; most of this is due to the length of a manually implemented basic (i.e. direct AST rewriting) transformation that inserts ghost steps to rewrite resources after inlining the tree reduction routine. We hope that with more proof automation to take care of some of these steps automatically, the script can be made more concise.

# 8 Related Work

*High-level GPU Languages.* Halide [20] and TVM [9] popularized the notion of using scheduling directives to generate high-performance imperative code, including for GPUs, starting from pure functional programs operating on arrays. Broadly speaking, these scheduling directives allow the user to specify when and where data is computed, capturing the most salient aspects of generating high-performance imperative code. RISE/ELEVATE [10] also followed in this paradigm, proposing improvements to the scheduling abstraction used by Halide and TVM. Futhark [12], Accelerate [8], and MDH [21] are other functional languages with compilers to GPUs, although they use automatic compilation instead of being driven by scheduling directives. All of these languages provide guarantees against data races and deadlocks by construction, as they abstract away from their input the low-level details where these bugs occur.

*Imperative, Safe GPU languages.* Another line of work has focused on type systems and language abstractions to ensure safety on lower-level, imperative GPU code, exposing a more direct mapping to CUDA that involves minimal performance decisions in the code generation phase. Descend [13] introduces a language and type system with Rust-like [22] borrow checking to track memory ownership on threads, thereby guaranteeing against data races. Its main limitation is its reliance on a set of pre-defined safe access patterns that map thread IDs to the array indices they own (like perfect tiling, or accessing in reverse, stride, etc.), outside of which an `unsafe` trap must be used, where the typechecker no longer ensures safety. Descend also models GPU parallelism explicitly with the loop-like `sched` construct, while making barrier placement explicit instead of relying on fork/join semantics. The `sched` construct has served as an inspiration for OptiGPU's `thread for`.

Prism [3] enforces correct execution of collective GPU operations, which include barriers but also modern GPU features like tensor core instructions and asynchronous copies, on an imperative language similar in abstraction to Descend's. It guarantees against deadlocks as a result, but provides looser guarantees around the lack of data races. It introduces the concept of *perspectives*, which track the threads executing a given piece of code at the type system level. Descend has a similar concept that is also used to prevent deadlocks, but perspectives are more generalized, allowing Prism to support newer GPU features. Perspectives have served as an inspiration for the execution contexts we have added as permissions in OptiGPU.

*Functional verification for GPUs.* Prior work has investigated functional verification, program logics, and formal semantics for GPU programs. Asakura et. al [1] prove soundness of a variant of concurrent separation logic for a minimal GPU language in Rocq, supporting threads and block-level barriers but no shared memory or host-side code. Alpinist [23, 24] applies the proof-preserving compilation technique to GPUs, which they call *annotation-aware transformations*. These annotations consist of loop transformations like tiling and fusion, similar to those supported by OptiGPU. The input to Alpinist is a verified GPU program (with host code) in separation logic (using the VerCors [6] framework) and a set of these transformations, which derive a verified, optimized GPU program. The primary difference between Alpinist and OptiGPU is that we start with a CPU program and capture the GPU refinement/lowering process within proof-preserving transformations, whereas Alpinist expects GPU code as input to begin with. In addition, Alpinist does not currently support shared memory.

Most recently, Kuiper [15] extends the existing PulseCore separation logic verification framework with GPU support. Kuiper allows functional verification, with guarantees against data race freedom, of GPU programs written in a low-level imperative language that aims to replicate CUDA as close as possible. It does not use any loop-like abstractions for parallelism and instead uses the same kernel function and built-in index approach that CUDA does. Kuiper supports a standard CUDA featureset like OptiGPU, as well as modern features like tensor cores and asynchrony. That said, it currently does not model collective execution in its logic, meaning they cannot guarantee against deadlocks due to thread divergence.

Barriers, tensor core instructions, etc. are viewed as single-thread operations that carry side proof obligations to ensure safe use of memory resources.

# 9 Conclusion and Future Work

We presented OptiGPU, a system for safe, verified, and low-level GPU programming that uses proof-preserving compilation to reduce proof burden on users. OptiGPU is implemented as extensions to the language, program logic, and transformations of an existing proof-preserving compilation framework for CPUs called OptiTrust. OptiGPU allows users to derive optimized, verified GPU code that can be generated to CUDA, starting from a naive verified CPU implementation. Thanks to reinterpreting the GPU programming model around parallel loop nests, most of this derivation happens while the program is still CPU-only, using only existing OptiTrust transformations. This allows OptiGPU users to concentrate on the structure of their GPU program – on both the host and device – including the scheduling of threads and data movement, before having to worry about parallelism, inserting barriers, or the exact GPU instructions and memories used in the final program. OptiGPU's extensions are then focused on those final, GPU-specific steps in the derivation process.

We implemented 2 case studies highlighting OptiGPU's ability to express and ensure safety for essential features of real-world CUDA programs, notably shared memory and thread divergence-free barriers. OptiGPU's performance on these case studies currently falls on average 13.7% short of the corresponding handwritten reference CUDA. We plan to address the unoptimized code generation and missing transformations that we have tested to be the cause for this performance discrepancy.

In addition, we hope to expand OptiGPU to more case studies. We have started a matrix multiplication case study and completed much of the CPU transformation portion, but have not yet applied GPU conversion transformations and generated CUDA. In future work, we would like to carry out case studies for other common GPU programs like stencils, parallel prefix sum, and histograms. Expanding OptiGPU to support GPU features like tensor cores and asynchrony is another direction we are hopeful to address.

Finally, we would like to investigate an operational semantics for the GPU language presented in Chapter 4; doing so would allow us to prove the soundness of our logic extensions for GPUs, and formalize our guarantees of data race freedom and deadlock freedom on OptiGPU code.